\documentclass[a4paper,twocolumn,superscriptaddress,8pt,
accepted=2017-09-12]{quantumarticle}
\usepackage[numbers,sort&compress]{natbib}
\usepackage{graphicx}
\usepackage{amsmath}
\usepackage{amssymb}

\usepackage{color}

\usepackage[english]{babel}
\usepackage{subfig}
\usepackage[T1]{fontenc}
\usepackage{mathrsfs}
\usepackage[colorlinks]{hyperref}

\newcommand{\pton}[1]{\left(#1\right)}
\newcommand{\pqua}[1]{\left[#1\right]}
\newcommand{\comm}[2]{\left[#1,#2\right]}
\newcommand{\der}[2]{\frac{\partial #1}{\partial #2}}

\newcommand{\md}[1]{\left|#1\right|}
\newcommand{\bk}[1]{\textbf{#1}}
\newcommand{\ave}[1]{\langle #1 \rangle}

\begin{document}

\title{Bose polaron as an instance of quantum Brownian motion}

\author{Aniello Lampo}\email{aniello.lampo@icfo.eu}
        \affiliation{ICFO -- Institut de Ciencies Fotoniques, The Barcelona Institute of Science and Technology, 08860 Castelldefels (Barcelona), Spain}
\author{Soon Hoe Lim}
        \affiliation{Department of Mathematics (Program in Applied Mathematics), University of Arizona, Tucson, AZ 85721-0089, USA}
\author{Miguel \'{A}ngel Garc\'{i}a-March}
        \affiliation{ICFO -- Institut de Ciencies Fotoniques, The Barcelona Institute of Science and Technology, 08860 Castelldefels (Barcelona), Spain}
\author{Maciej Lewenstein}
        \affiliation{ICFO -- Institut de Ciencies Fotoniques, The Barcelona Institute of Science and Technology, 08860 Castelldefels (Barcelona), Spain}
        \affiliation{ICREA -- Instituci{\'o} Catalana de Recerca i Estudis Avan\c{c}ats, Lluis Companys 23, E-08010 Barcelona, Spain}

\begin{abstract}
We study the dynamics of a quantum impurity immersed in a Bose-Einstein condensate as an open quantum system in the framework of the quantum Brownian motion model.  We derive a generalized Langevin equation for the position of the impurity. The Langevin equation is an integro-differential equation that contains a memory kernel and is driven by a colored noise. These result from considering the environment as given by the degrees of freedom of the quantum gas, and thus depend on its parameters, e.g. interaction strength between the bosons, temperature, etc. We study the role of the memory on the dynamics of the impurity. When the impurity is untrapped, we find that it exhibits a super-diffusive behavior at long times.  We find that back-flow in energy between the environment and the impurity occurs during evolution.  When the particle is trapped, we calculate the variance of the position and momentum to determine how they compare with the Heisenberg limit. One important result of this paper is that we find position squeezing for the trapped impurity at long times.   We  determine the regime of validity of our model and  the parameters in which these effects can be observed  in realistic experiments. 
\end{abstract}

\maketitle

\section{Introduction}

The concept of polaron has been introduced by Landau and Pekar  to describe the behavior of an electron  in a dielectric crystal~\cite{Landau48}. The motion of the electron distorts the spatial configuration of the surrounding ions,  which let their equilibrium positions to screen its charge. 
The movement of the ions is associated to phonon excitations that dress the electron. The resulting system, which consists of the electron and its surrounding phonon cloud, is called a polaron.  
The concept of polaron has been extended to describe a generic particle, the impurity, in a generic material, e.g.  a conductor, a semiconductor or a gas~\cite{Frolich1954,Alexandrov2009}.    
One important example is that of an impurity embedded in an ultracold gas. This system has been widely studied both theoretically and experimentally, in the case of a ultracold Fermi~\cite{Schirotzek2009,Kohstall2012,Koschorreck2012,
MassignanPolRev2014,Lan2014,Levinsen2014,Schmidt2012} or Bose gas~\cite{Cote2002,Massignan2005,Cucchietti2006,Palzer2009,Catani2012,
Spethmann2012,Rath2013,Fukuhara2013,
Shashi2014,Benjamin2014,Grusdt2014a,Grusdt2014b,Chris2015,Levinsen2015,Ardila2015,Volosniev2015,Grusdt2016,Grusdt2016Feb,
Shchadilova2016,Shchadilova2016b,Castelnovo2016,Ardila2016,Robinson2016,Jor2016,Hu2016,Rentrop2016}.
We focus on the latter case, where the Bose gas hosting the impurity is a Bose-Einstein condensate (BEC). 
 
In this paper, we study the physics of the impurity as an open system in the framework of quantum Brownian motion (QBM) model. In general,  the QBM describes a Brownian particle moving in a thermal bath consisting of a collection of non-interacting harmonic oscillators satisfying the Bose-Einstein statistics ~\cite{GardinerBook,BreuerBook,SchlosshauerBook, Schlosshauer2005,Zurek2003,Caldeira1983a,Caldeira1983b}. 
In our context, the impurity  plays the role of the Brownian particle, while the bath consists of the Bogoliubov modes of the BEC. 
We assume that the bath is homogeneous, i.e. the density of the BEC is space-independent. 

The open quantum system approach has been used recently in the context of ultracold quantum gases. For instance, the system of a bright soliton in a superfluid in one dimension in~\cite{Efimikin2013},  the system of a dark soliton in a one-dimensional BEC coupled to a non-interacting Fermi gas in~\cite{Hurst2016}, the system of the component of a moving superfluid \cite{Keser2016}, and the system of an impurity in a Luttinger liquid in~\cite{Cugliandolo2012, Bonart2013}.



There are a few formalisms within the open system approach that one can adopt to study the dynamics of the system. One of them is the master equation which describes the evolution of the reduced density matrix of the system.   
A recent study of the master equation for QBM has been performed in~\cite{Massignan2015}, both for linear and nonlinear coupling. 
There the master equation was derived using the Born-Markov approximation. Therefore, one possible way to study  the dynamics of the impurity in both homogeneous and inhomogeneous BEC is to use such master equations. 
However, the Born-Markov approximation is only valid at high temperature and for sufficiently weak coupling between the environment and the system. 
In fact, as we go beyond these restrictions we detect a breakdown of the positivity of the solution, corresponding to a violation of the Heisenberg uncertainty principle in these regimes.
One remedy to fix these problems is to cast the master equation in a Lindblad form by adding terms that are negligible in these regimes to the Born-Markov master equation.  
In~\cite{Lampo2016} a Lindblad model for QBM has been studied, for both  linear and nonlinear coupling. 
This allows one to explore the low temperature regime, as well as to consider values of the coupling between the environment and the system stronger than those permitted by the Born-Markov treatment. The drawback to this remedy is that the resulting Lindblad equation cannot be derived directly from the microscopic Hamiltonian of the system. Therefore, it cannot be adopted to investigate the dynamics of the impurity.

In view of the above remarks, we employ an alternative formalism to study our system: the Heisenberg equations for observables. 
This formalism not only permits us to go beyond the approximations underlying the works in~\cite{Massignan2015, Lampo2016}, but  is also simpler compared to the master equation formalism, since deriving an exact master equation for QBM is more difficult analytically.   
Moreover, as the main goal of the paper is to study the evolution of the observables of the impurity, Heisenberg formalism is particularly well suited for this task.   


We first show that the Hamiltonian of an impurity in an ultracold gas can be cast as a Hamiltonian of QBM model. We find that the coupling between the  system and the environment in the resulting Hamiltonian is nonlinear. A key assumption in this work and previous approaches~\cite{Efimikin2013,Cugliandolo2012} is to approximate this coupling by a linear one. Here we present a detailed study of the regimes of parameters in which this assumption is valid, and evaluate them in view of current experimental  feasibility for this system. In particular, we find the regimes of temperatures, interaction strength, experimental time and trapping frequency for the impurity in which the model is valid. We derive the spectral density (SD) of the system, which encapsulates  the effect of the bath on the central system. The spectral density is a super-ohmic one, corresponding to the presence of memory effects in the dynamics of the system. We then derive quantum Langevin equations to describe the evolution of relevant observables associated to the impurity. We solve these equations analytically whenever possible and numerically  for both configurations of the impurity. In the configuration where the impurity is untrapped, we calculate the mean square displacement, a measurable quantity in experiments~\cite{Catani2012}. We find that the particle exhibits a super-diffusive behavior, with a quadratic in time mean square displacement at long times. We show that a  back-flow of energy between the system and the environment occurs during evolution. When the impurity is trapped, we calculate its  position and momentum variance, analyzing its dependence on  system parameters such as temperature and coupling strength.  We detect new effects concerning the structure of the uncertainties ellipse:  genuine position squeezing occurs in the system, corresponding to high spatial localization of the impurity. This is a very relevant result of this paper.  We find the regime of experimentally feasible parameters to optimize such squeezing.  

The manuscript is organized as follows.  
In Sec.~\ref{HamSec} we present the Hamiltonian of an impurity in an ultracold gas,  discuss the steps to derive a QBM model for the system and justify the linear approximation for the coupling. In  Sec.~\ref{SDSec} we derive the SD of the system. In Sec.\ \ref{HeisenbergSec} we derive the quantum Langevin equations, and solve them  in Secs.\ \ref{UntrappedSec} and \ref{TrappedSec} for the untrapped and trapped configuration, respectively. In the Appendices, we provide the details of the derivations, analysis of the feasibility of the approach and of the presented results in the parameter regimes.

\section{Hamiltonian}\label{HamSec}
We consider a single impurity atom with mass $m_{\rm I}$ immersed in a $d-$dimensional gas of $N$ bosons with mass $m_{\rm B}$.
The interactions among the bosons are described by a potential $V_{\rm B}(\bk{x})$.  Let $\Psi(\bk{x})$,  $\Psi^\dag(\bk{x})$ denote the annihilation and creation operators of atoms at the position $\bk{x}$. They fulfill the canonical bosonic commutation relation, $[ \Psi(\bk{x}),  \Psi^\dag(\bk{x'})]=\delta(\bk{x} -\bk{x'})$. Bosonic density thus reads $n_{\rm B}=\Psi^\dag(\bk{x})\Psi(\bk{x})$. 
The Hamiltonian of the system is 
\begin{equation}\label{Ham1}
H=H_{\rm I}+H_{\rm B}+H_{\rm BB}+H_{\rm IB},
\end{equation}
where:
\begin{align}
&H_{\rm I}=\frac{{\bk{p}}^2}{2m_{\rm I}}+U(\bk{x}),\label{H_I}\\
&H_{\rm B}=\int d^d \bk{x} \,\Psi^{\dag}(\bk{x})\left(\frac{\bk{p}_{\rm B}^2}{2m_{\rm B}}+V(\bk{x})\right)\Psi(\bk{x})\\
&=\sum_{\bk{k}}\epsilon_{\rm \bk{k}}
{a}^{\dagger}_{\bk{k}}
{a}_{\bk{k}},\label{H_B}\\
&H_{\rm BB}=g_{\rm B}\int d^d\bk{x}\, \Psi^\dag(\bk{x})\Psi^\dag(\bk{x})\Psi(\bk{x}) \Psi(\bk{x})\\
&=\frac{1}{2V}\sum_{\bk{k},\bk{k'},\bk{q}}V_{\rm B}(\bk{q})
{a}^{\dagger}_{\bk{k'}-\bk{q}}
{a}^{\dagger}_{\bk{k}+\bk{q}}
{a}_{\bk{k'}}
{a}_{\bk{k}}\label{H_BB},\\
&H_{\rm IB}= g_{\rm IB} n_{\rm B}(\bk{x})\\
&=\frac{1}{V}\sum_{\bk{k},\bk{q}}V_{\rm IB}(\bk{k})
{\rho}_{\rm I}(\bk{q})
{a}^{\dagger}_{\bk{k}-\bk{q}}
{a}_{\bk{k}}\label{H_IB},
\end{align}
in which the operator $a_{\bk{k}}$ ($a^{\dagger}_{\bk{k}}$) destroys (creates) a boson of mass $m_{\rm B}$, wave vector $\bk{k}$, and energy $\epsilon_{\bk{k}}=(\hbar k)^2/(2m_{\rm B})-\mu$, measured from its chemical potential, $\mu$. 
Equation~(\ref{H_I}) is the Hamiltonian for the free impurity.   Equation~(\ref{H_B}) is the  Hamiltonian of non-interacting bosons in a potential $V(\bk{x})$. We consider in the following that the potential is homogeneous and the system is  enclosed in a box of volume $V$. The last two terms in Eqs.~(\ref{H_BB}) and (\ref{H_IB}) are the interaction among the atoms of the gas and between them and the impurity, respectively. 
The quantities 
\begin{align}
&V_{\rm B}(\bk{q})=\mathcal{F}_{\bk{q}}\pqua{g_{\rm B}\delta(\bk{x}-\bk{x'})},\\ 
&V_{\rm IB}(\bk{q})=\mathcal{F}_{\bk{q}}\pqua{g_{\rm IB}\delta(\bk{x}-\bk{x'})},
\end{align}
represent respectively the Fourier transforms of the boson-boson and impurity-boson interaction, with
\begin{equation}
g_{\rm B}=4\pi\hbar^2a_{\rm B}/m_{\rm B},\quad g_{\rm IB}=2\pi\hbar^2a_{\rm IB}/m_{\rm R}.
\end{equation}
Here $a_{\rm B}$ is the scattering length between two identical bosons and $a_{\rm IB}$ represents that between the impurity and the BEC bosons.  The reduced mass is $m_{\rm R}=m_{\rm B} m_{\rm I}/(m_{\rm B}+m_{\rm I})$ and the (dimensionless) density of the impurity in the momentum domain is
\begin{equation}
{\rho}_{\rm I}(\bk{q})=\int^{+\infty}_{-\infty}e^{-i\bk{q}\cdot\tilde{\bk{x}}}\delta\pton{\tilde{\bk{x}}-
\bk{x}}d^3\tilde{\bk{x}}.
\end{equation}
In experiments, we usually have more than one impurity in the gas, so one can include a term modeling the interaction between several impurities in  Eq.~$\pton{\ref{Ham1}}$. 
Here we consider that the impurities concentration is low enough as to neglect such an additional interacting term. It is important to realize that the Hamiltonian is positively defined, and as such cannot lead to instabilities -- this is clearly seen from the form of the various parts of the Hamiltonian in the position representation. 

To obtain the QBM form of the Hamiltonian we first replace the creation/annihilation operator in the fundamental state by its average value $\sqrt{N_{0}}$.
For bosons below the critical temperature  the atoms mainly occupy the ground mode, with negligible fluctuations to other modes, thus forming a BEC.
Consequently, we neglect the terms proportional to $N_{\bk{k}}$ $(\bk{k}\neq\bk{0})$, i.e. the number of particles out of the ground state. 
Then, we apply the Bogoliubov transformation
\begin{equation}
{a}_{\bk{k}}=u_{\bk{k}}
{b}_{\bk{k}}-v_{\bk{k}}
{b}_{-\bk{k}}^{\dagger},\quad
{a}_{-\bk{k}}=u_{\bk{k}}
{b}_{-\bk{k}}-v_{\bk{k}}
{b}_{\bk{k}}^{\dagger}\label{BT},
\end{equation}
with
\begin{align}
&u^{2}_{\bk{k}}=\frac{1}{2}\pton{\frac{\epsilon_{\bk{k}}+n_0V_{\rm B}}{E_{\bk{k}}}+1},\\ 
&v^{2}_{\bk{k}}=\frac{1}{2}\pton{\frac{\epsilon_{\bk{k}}+n_0V_{\rm B}}{E_{\bk{k}}}-1},
\end{align}
in which
\begin{equation}\label{Bog}
E_{\bk{k}}=\hbar c|\bk{k}|\sqrt{1+\frac{1}{2}\pton{\xi\bk{k}}^2}\equiv\hbar\omega_{\bk{k}}
\end{equation}
is the Bogoliubov spectrum and $n_{\rm 0}$ is the density of particles in the ground state.
Since we consider a homogeneous gas, the density $n_{\rm 0}$ is constant. 
In Eq.~(\ref{Bog}) the quantities
\begin{equation}
\xi=\frac{\hbar}{\sqrt{2 g_{\rm B} m_{\rm B} n_{\rm 0}}},\quad c=\sqrt{\frac{g_{\rm B}n_{\rm 0}}{m_{\rm B}}}=\frac{\hbar}{\sqrt{2}m_{\rm B}\xi}
\end{equation} 
represent respectively the coherence length and the speed of sound. 

The transformations~(\ref{BT}) diagonalize the terms describing the condensed atoms (see e.g.~\cite{Pita-String})
\begin{equation}\label{HamAftBog}
{H}_{\rm B}+H_{\rm BB}=\sum_{\bk{k}\neq\bk{0}}E_{\bk{k}}
{b}^{\dagger}_{\bk{k}}
{b}_{\bk{k}},
\end{equation}
apart of a few non-operator terms, simply providing a shift of the energy levels of the atoms of the BEC.

We treat in the same manner the bosons-impurity interaction.   We are going now to keep only the terms proportional to $\sqrt{N_{0}}$ -- this is in principle a well motivated approximation, since  the condensate is macroscopically occupied and  $N_{i\neq0}\ll N_0$. Unfortunately, this approximation  is dangerous. Our model generically has an ultraviolet divergence, like most of models of the non-relativistic quantum field theory. We do all the calculations with a physical cut-off, so that the ultraviolet divergences do not really affect our theory. Still, at large values of the cut-off we might expect large negative shifts of the impurity energy, that might cause unphysical instability. After dropping out the terms bilinear in   $N_{i\neq0}$ we obtain
\begin{equation}
{H}_{\rm IB}=n_{0}V_{\rm IB}+\sqrt{\frac{n_{\rm 0}}{V}}\sum_{\bk{k}\neq\bk{0}}
{\rho}_{\rm I}(\bk{k})V_{\rm IB}\pton{
{a}_{\bk{k}}+
{a}^{\dagger}_{-\bk{k}}}\label{HIB1}.
\end{equation}
The first term on the right hand-side represents the mean field energy. 
It is a constant just providing a shift of the energy of the polaron, so we will neglect it in what follows.
Inserting Eq.\ \eqref{BT} into Eq.\ \eqref{HIB1}, one gets
\begin{align}
{H}_{\rm IB}=&\sqrt{\frac{n_{\rm 0}}{V}}\sum_{\bk{k}\neq\bk{0}}
{\rho}_{\rm I}(\bk{k})V_{\rm IB}\pton{u_\bk{k}-v_\bk{k}}\pton{{b}_{\bk{k}}+
{b}^{\dagger}_{-\bk{k}}}\label{HIB2}\\ \nonumber
=&\sqrt{\frac{n_{\rm 0}}{V}}\sum_{\bk{k}\neq\bk{0}}
{\rho}_{\rm I}(\bk{k})V_{\rm IB}\sqrt{\frac{\epsilon_{\bk{k}}}{E_{\bk{k}}}}\pton{
{b}_{\bk{k}}+
{b}^{\dagger}_{-\bk{k}}},
\end{align}
where we again neglected the non-operator terms. 
\\By some algebra, the expression in Eq.~(\ref{HIB2}) reads 
\begin{equation}\label{IntTerm}
{H}_{\rm IB}=\sum_{\bk{k}\neq\bk{0}}V_{\bk{k}}e^{i\bk{k}\cdot
{\bk{x}}}\pton{
{b}_{\bk{k}}+
{b}^{\dagger}_{-\bk{k}}},
\end{equation}
in which
\begin{equation}
V_{\bk{k}}=g_{\rm IB}\sqrt{\frac{n_{\rm 0}}{V}}\pqua{\frac{(\xi k)^2}{(\xi k)^2+2}}^{\frac{1}{4}}.
\end{equation}
The expression in Eq.\ \eqref{IntTerm} is known in the literature as the interaction term of the Fr\"{o}hlich Hamiltonian. 
We restrict to the limit
\begin{equation}\label{mainAssumption}
\textbf{k}\cdot \textbf{x}\ll1. 
\end{equation}
  Accordingly Eq.~(\ref{IntTerm}) assumes the form:
\begin{equation}\label{HIBExp1Ord}
{H}_{\rm IB}=\sum_{\bk{k}\neq\bk{0}}V_{\bk{k}}\left(\mathbb{I}+i\bk{k}\cdot\bk{x}\right)\pton{
{b}_{\bk{k}}+
{b}^{\dagger}_{-\bk{k}}}. 
\end{equation}
Equation~$\pton{\ref{mainAssumption}} $ is a crucial assumption of this paper. It allows for a linear coupling between the impurities and the bosons, which will play the role of an environment. This is crucial when considering this system from an open quantum system perspective. In Appendix~\ref{ValidityApp} we justify when this assumption is valid as a function of the relevant physical parameters in the problem.
A similar assumption is considered in~\cite{Efimikin2013} to study the physics of a bright soliton in a superfluid, and in~\cite{Cugliandolo2012} where QBM has been employed to treat the dynamics of an impurity in a Luttinger Liquid.

The resulting Hamiltonian of the impurity in a BEC is
\begin{equation}
{H}=H_{\rm I}+\sum_{\bk{k}\neq\bk{0}}E_{\bk{k}}
{b}^{\dagger}_{\bk{k}}
{b}_{\bk{k}}+\sum_{\bk{k}\neq\bk{0}}\hbar g_{\bk{k}}\pi_{\bk{k}}\bk{x}\label{HfinLin},
\end{equation}
with
\begin{equation}\label{DimensionlessMomenta}
g_{\bk{k}}=\bk{k} V_{\bk{k}}/\hbar,\quad {\pi}_{\bk{k}}=i\pton{ {b}_{\bk{k}}- {b}^{\dagger}_{\bk{k}}}.
\end{equation} 
To get Eq.~(\ref{HfinLin}) we redefined the Bogoliubov modes operators as
$
{b}_{\bk{k}}\rightarrow {b}_{\bk{k}}-V_{\bk{k}}/E_{\bk{k}}\mathbb{I},
$
in order to get rid of the term in Eq.~(\ref{IntTerm}) proportional to the identity operator. 
This operation yields a non-operator term which has been neglected in agreement with the procedure above. 

The Hamiltonian in Eq.~(\ref{HfinLin}) describes an impurity coupled to a bath of Bogoliubov modes through an interaction term linearly dependent on the impurity position. This is exactly the same situation of the QBM. Here the impurity plays the role of the Brownian particle, while the Bogoliubov modes represents the environment. 
The Hamiltonian in Eq.~(\ref{HfinLin}) is in fact almost the same of that of the QBM model. The only difference lies in the dependence of the interaction term on the Bogoliubov modes operator: while in the QBM model it depends on their positions, for the present system it depends on their (dimensionless) momenta, $\pi_{\bk{k}}$. We will see in the following that the two situations are equivalent, and the theory of the QBM can be exploited to investigate the impurity problem. 
 
Unfortunately, the Hamiltonian in Eq.~(\ref{HfinLin}) is not positively defined
(note that at this point we match the situation described in \cite{Sanchez1994}, where translational symmetry is broken).
We could repair this by taking into account bilinear terms in  ${a}_{\bk{k}}$ and ${a}^{\dagger}_{\bk{k}}$  in Eq.~(\ref{HIB1}), or equivalently ${b}_{\bk{k}}$ and ${b}^{\dagger}_{\bk{k}}$, as in \cite{Bruderer2007,Rath2013,Brunn2015,Shchadilova2016,Shchadilova2016b}. 
One way of curing it is to include these terms in the theory in an exact manner, which is possible, but technically complex.  It is  much easier to use the same trick as Caldeira and Leggett used in their seminal paper -- i.e. complete the Hamiltonian to a positively defined one by writing
\begin{equation}
{H}=H_{\rm I}+\sum_{\bk{k}\neq\bk{0}}E_{\bk{k}}
\left({b}^{\dagger}_{\bk{k}}+i \hbar g_{\bk{k}} \bk{x}/E_{\bk{k}}\right)\left(
{b}_{\bk{k}}-i\hbar g_{\bk{k}} \bk{x}/E_{\bk{k}}\right)\label{HfinLinReno}.
\end{equation}
Clearly, Caldeira-Leggett remedy leads directly to the trapping harmonic potential for the impurity that cancels the negative harmonic frequency shift that appears in the absence of the compensation term. 
In Appendix~\ref{DerHE} it is shown how this negative contribution arises by evaluating its effect directly in the equations of motion derived from Hamiltonian~(\ref{HfinLin}). The equations of motion are presented in Sec.~\ref{HeisenbergSec}. 

The Hamiltonian in Eq.\ \eqref{HfinLin} is the first step to put the Bose polaron problem in the framework of quantum Brownian motion. 
This result has been obtained by considering the Hamiltonian in Eq.\ \eqref{Ham1}, which is a conventional choice in the context of polaron physics and it is largely used in the literature, for example in \cite{Tempere2009, Shashi2014, Casteels2014}.
Nevertheless, this Hamiltonian is not fully appropriate if we push our analysis towards the strong coupling regime.
Here, the interaction term in Eq.~(\ref{H_IB}) needs to be generalized by including a quadratic dependence on the operators of the impurity.
Accordingly, when one introduces Bogoliubov operators, the interaction term in Eq.\ \eqref{IntTerm} includes additional terms manifesting a quadratic, rather than linear, dependence on these operators. In this way one goes beyond the Fr\"{o}hlich paradigm. 
Such a generalization is widely studied nowadays, for instance in~\cite{Bruderer2007,Rath2013,Brunn2015,Shchadilova2016,Shchadilova2016b}.
In particular there is still an open debate concerning the validity regime of the Fr\"{o}hlich Hamiltonian, i.e. for which values of the system parameters the quadratic Bogoliubov operators terms can be dropped out. 
In the present paper we neglect these terms, looking to the traditional Fr\"{o}hlich interaction term in Eq.\ \eqref{IntTerm}.
Of course this choice does not allow us to explore the strong coupling regime, but we shall explain in the following that it is appropriate for the values of the system parameters we consider.

\section{Spectral Density}\label{SDSec}

The last term of Hamiltonian~(\ref{HfinLin}) is a coupling between the  impurity position and  the momenta  of the bath oscillators, $\pi_{\bk{k}}$. This plays the role of the interaction Hamiltonian between the system and the bath oscillators, when compared with the QBM Hamiltonian.  The difference is that the coupling occurs with the momenta of the oscillators  rather than on their positions. In the following, we show  that both situations are  equivalent.   
In fact,  the interaction with the Bogoliubov modes enters in the dynamics of the system only through the self-correlation function of the environment, defined as
\begin{equation}
\mathcal{C}(\tau)=\sum_{\bk{k}\neq\bk{0}}\hbar g^2_{\bk{k}}\ave{ {\pi}_{\bk{k}}(\tau) {\pi}_{\bk{k}}(0)}.
\end{equation}
Replacing the expression for the dimensionless momenta $\pi_{\bk{k}}$ in Eq.~(\ref{DimensionlessMomenta}) and recalling that the environment is made up by bosons,
\begin{equation}
\ave{ {b}^\dagger_{\bk{k}} {b}_{\bk{k}}}=\frac{1}{\exp\left(\hbar\omega_{\bk{k}}/k_{\rm B}T\right)-1},
\end{equation}
we find
\begin{align}
\mathcal{C}(\tau)
=&\sum_{\bk{k}\neq\bk{0}}\hbar g^2_{\bk{k}}\left[\coth\left(\frac{\hbar\omega_{\bk{k}}}{2k_{\rm B}T}\right)\cos\left(\omega_{\bk{k}}\tau\right)-i\sin\left(\omega_{\bk{k}}\tau\right)\right]\\\equiv&\nu(\tau)-i\lambda(\tau),
\end{align}
where
\begin{align}
&\nu(\tau)=\int^{\infty}_{0}J(\omega)\coth\left(\frac{\hbar\omega}{2k_{\rm B}T}\right)\cos\left(\omega\tau\right)d\omega,\label{NoiseKernel}\\
&\lambda(\tau)=\int^{\infty}_{0}J(\omega)\sin\left(\omega\tau\right)d\omega=-m_{\rm I}\dot{\Gamma}(t),\label{dampingKernel}
\end{align}
are respectively the noise and dissipation kernel, representing the real and imaginary part of the self-correlation function.
The latter is related to the damping kernel, which is defined as
\begin{equation}\label{DampingKernel}
\Gamma(\tau)=\frac{1}{m_{\rm I}}\int^{\infty}_0\frac{J(\omega)}{\omega}\cos(\omega\tau)d\omega. 
\end{equation}
A crucial function in the expressions above is the {\it spectral density } (SD) 
\begin{equation}\label{SDdef}
J(\omega)=\sum_{\bk{k}\neq\bk{0}}\hbar g^2_{\bk{k}}\delta\pton{\omega-\omega_{\bk{k}}}.
\end{equation}
The SD plays a fundamental role on characterizing the dynamics of QBM. It contains the information about the environment after averaging over its degrees of freedom. 
Note that this quantity depends on the square of the coupling constant. 
This is the reason why, as we will see in the following, our theory does not depend on the sign of the interaction.

The rest of this section is devoted to the derivation of the SD.
We first assume that the environment is large, that is,  the large number of oscillators within it allows to switch from a discrete to a continuous distribution of Bogoliubov modes in the frequency domain. Then,  in the definition of the SD, Eq.\ (\ref{SDdef}), we turn the discrete sum into an integral,
\begin{equation}
\sum_{\bk{k}}\rightarrow\int\frac{V}{(2\pi)^{\rm d}}d^{\rm d}k.
\end{equation}
Using the relation
\begin{equation}
\delta\pton{\omega-\omega_{\bk{k}}}=\frac{1}{\partial_{\bk{k}}\omega_{\bk{k}}|_{\bk{k}=\bk{k}_{\omega}}}\delta\pton{\bk{k}-\bk{k}_{\omega}},
\end{equation}
one finds
\begin{equation}\label{SDdef1}
J(\omega)\!=\!\frac{n_{\rm 0}g^2_{\rm IB}}{\hbar}\frac{S_{\rm d}}{(2\pi)^{\text{d}}}\int dkk^{\rm d+1}\sqrt{\frac{(\xi\bk{k})^2}{(\xi\bk{k})^2+2}} \frac{\delta\pton{\bk{k}-\bk{k}_{\omega}}}{\partial_{\bk{k}}\omega_{\bk{k}}|_{\bk{k}=\bk{k}_{\omega}}},
\end{equation}
where
\begin{equation}\label{InvBog}
 k_\omega=\xi^{-1}\sqrt{\sqrt{1+2\pton{\xi\omega/c}^2}-1},
\end{equation}
is the inverse of the Bogoliubov spectrum in Eq.~(\ref{Bog}). 
The quantity $S_{\rm d}$ is the surface of the hypersphere in the momentum space with radius $k$ in $\rm{d}-$dimensions. For $\rm{d}=1,2,3$ it reads
\begin{equation}
S_1=2,\quad S_2=2\pi,\quad S_3=4\pi^2.
\end{equation}
Hereafter, we focus on $\text{d}=1$, but the generalization to higher dimensions is conceptually immediate. 
Thus, in one dimension the SD is
\begin{align}\label{SD1D}
&J_{\rm 1d}(\omega)=m_{\rm I}\tilde{\tau}\omega^3\chi_{\rm 1d}(\omega),\\ &\tilde{\tau}=\frac{\eta^2}{2\pi m_{\rm I}}\left(\frac{m_{\rm B}}{n_{\rm 0}g^{\frac{1}{3}}_{\rm B}}\right)^{\frac{3}{2}},\\
&\chi_{\rm 1d}(\omega)=2\sqrt{2}\left(\frac{\Lambda}{\omega}\right)^3\frac{\left[\sqrt{1+\frac{\omega^2}{\Lambda^2}}-1\right]^{\frac{3}{2}}}{\sqrt{1+\frac{\omega^2}{\Lambda^2}}}\label{chi1},
\end{align}
where $\tilde{\tau}$ represents a relaxation time scale. We introduced the interaction strength
\begin{equation}\label{eta}
\eta=g_{\rm IB}/g_{\rm B}, 
\end{equation}
which expresses the strength of the impurity-bosons interaction in units of the bosons-bosons one. 
The majority of our results are expressed as a function of such a parameter because in experiments with ultracold gases it can be tuned~\cite{Catani2012}.
This parameter is the crucial one to define the regime of validity of our theory. 
In fact, in Sec.\ \ref{HamSec} we precised that the form of the Hamiltonian we use, showing a linear, rather than quadratic, dependence on the Bose operators, does not work for strong coupling. 
The parameter in Eq.\ \eqref{eta} allows to quantify to which extent the coupling has to be weak.
In particular in \cite{Grusdt2016,Grusdt2017} it has been shown that the standard Fr\"{o}hlich Hamiltonian in Eq.\ \eqref{IntTerm} without quadratic terms in the Bose operators holds if
\begin{equation}
\eta\lesssim\eta_c\equiv\pi\sqrt{n_{\rm 0}\md{a_{\rm B}}}=\pi\sqrt{\frac{2n_{\rm 0}}{m_{\rm B}g_{\rm B}}}.
\end{equation}
This equation has been derived for a gas in one dimension, but in \cite{Grusdt2016} the corresponding $3\text{D}$ generalization is presented.  
We consider the same situation of \cite{Catani2012}, i.e. a gas of Rb atoms with
\begin{equation}\label{validity}
n_{\rm 0}=7\left(\mu \text{m}\right)^{-1},\quad g_{\rm B}=2.36\cdot10^{-37}\text{J}\cdot\text{m}. 
\end{equation}
Accordingly we obtain
\begin{equation}
\eta_c\approx7,
\end{equation}
providing an upper bound for the acceptable values for the coupling strength.

The quantity
\begin{equation}\label{CutOffFrequency}
\Lambda=n_0g_{\rm B}/\hbar
\end{equation}
is a characteristic frequency distinguishing the high-frequency domain from the low-frequency one.
The SD is ohmic when it depends linearly on the frequency of the oscillators of the environment. 
This is not the case of the physical system we are dealing with. 
For $\omega\ll\Lambda$ we find
\begin{equation}\label{CubicSD}
J_{1\rm d}\sim\omega^3,
\end{equation}
namely SD is proportional to the third power of the frequency of the Bogoliubov modes. In our case, the SD is super-ohmic.  
In higher dimensions, the SD is also super-ohmic, as in general
\begin{equation}
J_{\rm d}\sim\omega^{2+\text{d}}.
\end{equation}
The super-ohmic dependence in Eq.~(\ref{CubicSD}) has been found in~\cite{Cugliandolo2012, Peotta2013} for an impurity immersed in a  Luttinger liquid, and in~\cite{Efimikin2013} for a bright soliton in a superfluid. 
This behavior is associated to the the linear part of the Bogoliubov spectrum. 
Hereafter we shall focus on the former considering
\begin{equation}\label{SDcut}
J(\omega)=m_{\rm I}\tilde{\tau}\omega^3\theta(\omega-\Lambda),
\end{equation}
namely we introduce a sharp ultraviolet cut-off to regularize the SD at high-frequency. 
We emphasize that the results we will present are independent on this ultraviolet regularization. 
Indeed, we reproduced the calculation by introducing an exponential, rather than a sharp cut-off, and we obtained the same results (see Appendix \ref{DampingKernelAppendix}).  

However, apart of its analytical form, the cut-off plays a crucial role. 
The results we shall present in the following depend on its presence, because it allows us to focus on the low frequency portion of the spectral density, forgetting about the high frequency one. 
Such a way to proceed is absolutely fitting if one is interested in the long-time dynamics. 
In this context the cut-off frequency in Eq.\ \eqref{CutOffFrequency}
is not an artificial quantity, but it arises in a natural manner: it is the characteristic frequency distinguishing the phononic linear part of the Bogoliubov spectrum from the quadratic one. 
This is more clear if we express it in terms of the traditional parameters of the Bogoliubov spectrum:
\begin{equation}\label{Lambda1}
\Lambda=c/\left(\sqrt{2}\xi\right)
\end{equation}
Precisely, considering $\omega/\Lambda\ll1$, the inverse of the Bogoliubov spectrum in Eq.\ \eqref{InvBog} takes the following form:
\begin{equation}
k_\omega\sim\omega,
\end{equation} 
namely the Bogoliubov dispersion relation gets linear. 
Replacing it in Eq.\ \eqref{SDdef1} we obtain the cubic behavior in Eq.\ \eqref{SD1D}.

Finally, the negligibility of the high frequency part of the SD, i.e. the ultraviolet cut-off, arises naturally if one looks to the phonon linear part of the Bogoliubov spectrum, which is reasonable if we want to study to long-time dynamics. 
It is important to highlight that in such a regime our results do not depend on whether the cut-off is present or not, as we proved in the end of Appendix \ref{DampingKernelAppendix}.
The SD takes thus the polynomial cubic structure in Eq.\ \eqref{SD1D}.
Developing the theory of QBM for such a cubic SD is a central part of the current work. In the following section we shall show that such a super-ohmic behavior is related to memory effects.

\section{Heisenberg Equations}\label{HeisenbergSec}
In this section we derive the Heisenberg equations of motion.
For the sake of simplicity, we focus on the case where the BEC is confined in one dimension.  We consider that the impurity is trapped in a harmonic potential, i.e., 
\begin{equation}
U(x)=m_{\rm I}\Omega^2x^2/2\label{HarmonicPotential},
\end{equation}
which is a common set-up in ultracold atom systems. The Heisenberg equations are
\begin{align}
&\dot{x}(t)=\frac{i}{\hbar}\comm{ {H}}{ {x}(t)}=\frac{ {p}(t)}{m_{\rm I}},\label{EqX}\\
&\dot{p}(t)\!=\!\frac{i}{\hbar}\comm{ {H}}{ {p}(t)}=-m_{\rm I}\Omega^2 {x}(t)-\hbar\sum_kg_k\pi_k(t),\label{EqP}\\
&\dot{b}_k(t)
=\frac{i}{\hbar}\comm{ {H}}{ {b}_k(t)}=-i\omega_k {b}_k(t)-g_k {x}(t),\label{Eqb}\\
&\dot{ {b}}^\dagger_k(t)=\frac{i}{\hbar}\comm{ {H}}{ {b}^\dagger_k(t)}=+i\omega_k {b}^\dagger_k(t)-g_k {x}(t),\label{Eqbdag}
\end{align}
where the Hamiltonian is given by in Eq.~(\ref{HfinLin}). 
Equations~(\ref{EqX})-(\ref{Eqbdag}) can be combined to obtain
\begin{equation}\label{EqDiffFin}
\ddot{x}(t)+\Omega^2 x(t)+\der{}{t}\int^{t}_0\Gamma(t-s) x(s)ds=\frac{B(t)}{m_ {\rm I}},
\end{equation}
 as detailed in Appendix\ \ref{DerHE}. 
This is the equation of motion for the impurity position. 
It may be viewed as the quantum analogue of a classical stochastic differential equation, where 
\begin{equation} 
{B}(t)=\sum_k i\hbar g_{\rm k}( {b}^{\dagger}_{\rm k}e^{i\omega_{\rm k}t}- {b}_{\rm k}e^{-i\omega_{\rm k}t})
\end{equation}
plays the role of a stochastic force, and $\Gamma(t)$ is the damping kernel introduced in Eq.~(\ref{DampingKernel}), which expression may be obtained through an integration by parts:
\begin{equation}\label{DampKernelResult}
\Gamma(t)=\frac{\tilde{\tau}}{t^3}\left[2\Lambda t\cos\left(\Lambda   
 t\right)-(2-\Lambda^2t^2)\sin\left(\Lambda t\right)\right].
\end{equation}
An important feature of such an equation is that it is non-local in time, namely the temporal evolution of $x$ at time $t$ depends on  its past history, i.e. $x(s)$ with $s<t$. 
We can state that in general the dynamics of an impurity in a homogeneous BEC in one dimension, described by Eq.~(\ref{EqDiffFin}), carries a certain amount of \textit{memory effects}. 
Actually, there is one special situation where this equation reduces to a local one, and in particular to a traditional damped harmonic oscillator with a stochastic force. 
This is the case constituted by a damping kernel proportional to a Dirac delta. 
It is immediate to prove that this expression of the damping kernel results by a \textit{ohmic} SD, i.e. an SD depending linearly on the frequency. 
As we showed in Sec.~\ref{SDSec} this is not the case of the present system, where the SD is super-ohmic. 
In conclusion we can state that the dynamics of an impurity in a BEC always carries a certain amount of memory effects. 

Equation~(\ref{EqDiffFin}) is a second-order linear non-homogeneous differential equation. Its solution is the sum of a particular one and that of the related homogeneous equation. For the latter we can proceed by applying the Laplace transform operator, in order to obtain the solution in terms of the initial position and velocity. 
The particular solution, instead, is the convolution product of the Green function and the position operator.  
Therefore, the solution for the impurity position in the Heisenberg picture takes the following form
\begin{equation}\label{XHeis}
x(t)\!=\!G_1(t)x(0)+G_2(t)\dot{x}(0)+\frac{1}{m_{\rm I}}\int^{t}_0G_2(t-s)B(s)ds,
\end{equation} 
where the functions $G_1$ and $G_2$ are defined through their Laplace transforms 
\begin{align}
&\mathcal{L}_z[G_1(t)]=\frac{z+\mathcal{L}_z[\Gamma(t)]}{z^2+\Omega^2+z\mathcal{L}_z[\Gamma(t)]},\label{LTG1}\\
&\mathcal{L}_z[G_2(t)]=\frac{1}{z^2+\Omega^2+z\mathcal{L}_z[\Gamma(t)]},\label{LTG2}
\end{align}
and satisfy
\begin{align}
&G_1(0)=1,\quad \dot{G}_1(0)=0\label{InCondG1},\\
&G_2(0)=0,\quad \dot{G}_2(0)=1\label{InCondG2}.
\end{align}
Equations~(\ref{LTG1}) and~(\ref{LTG2}) depend on the Laplace transform of the damping kernel which is calculated in Appendix~\ref{DampingKernelAppendix}. Here we present only the final result
\begin{equation}\label{LTGamma}
\mathcal{L}_z\left[\Gamma(t)\right]=z\tilde{\tau}\left[\Lambda-z\arctan\left(\Lambda/z \right)\right].
\end{equation}
In order to find the expression of the position impurity as a function of the time, and so to characterize its motion, we have to invert the Laplace transforms in Eqs.~(\ref{LTG1}) and~(\ref{LTG2}).
In the following two sections, we will invert this transformations to obtain the complete solution both analytically and numerically, for the case of untrapped (Sec.~\ref{UntrappedSec}
) and trapped impurity (Sec.~\ref{TrappedSec}). 

\section{Untrapped Impurity}\label{UntrappedSec}
Here we consider an untrapped impurity, that is $\Omega=0$ in Eq.~(\ref{HarmonicPotential}).  
In this case the impurity position in the Heisenberg picture is described by Eq.~(\ref{XHeis}) with
\begin{align}
&\mathcal{L}_z[G_1(t)]=\frac{z+\mathcal{L}_z[\Gamma(t)]}{z^2+z\mathcal{L}_z[\Gamma(t)]}=1/z,\label{LTG1Un}\\
&\mathcal{L}_z[G_2(t)]=\frac{1}{z^2+z\mathcal{L}_z[\Gamma(t)]}.\label{LTG2Un}
\end{align}
To completely characterize the motion of the impurity one has to invert these Laplace transforms. 
For Eq.~(\ref{LTG1Un}) this calculation can be performed straightforwardly to get
\begin{equation}
G_1(t)=1.\label{G11}
\end{equation}
Thus the contribution of the initial position in Eq.~\eqref{XHeis} is constant in time. 
We highlight that such a result may be found even without knowledge of the form of the damping kernel, and therefore about the form of the SD. Thus, Eq.~(\ref{G11}) holds for any untrapped impurity, regardless of the details of the coupling. 

It is more difficult to handle with the Laplace transform in Eq.~(\ref{LTG2Un}), due to the arctangent in the Laplace transform of the damping kernel in Eq.~(\ref{LTGamma}). 
We are interested in the long-time regime, which corresponds to $z\ll\Lambda$. The asymptotic expansion of Eq.~(\ref{LTGamma}) in this limit reads
\begin{equation}
\mathcal{L}_z\left[\Gamma(t)\right]=z\tilde{\tau}\Lambda+o(z^2),
\end{equation}
and therefore
\begin{equation}
\mathcal{L}_z[G_2(t)]=\frac{1}{(1+\Lambda\tilde{\tau})z^2}\equiv\frac{1}{\alpha z^2},
\end{equation}
which can be inverted providing
\begin{equation}\label{G2LongTime}
G_2(t)=t/\alpha.
\end{equation}
We note that such an expression does not fulfill the boundary conditions in Eq.~\eqref{InCondG2}, in particular $\dot{G}_2(0)\neq1$, but this is justified by the fact that Eq.~\eqref{G2LongTime} refers to a long-time behavior. 

Equation~(\ref{G2LongTime}) has been obtained by expanding the Laplace transform of the damping kernel at the first order in $z/\Lambda$. 
In general one may perform such an expansion to the $n^{\text{th}}$ order, and the Laplace transform in Eq.~(\ref{LTG2Un}) takes the form of the inverse of a polynomial of $(n+1)^{\text{th}}$ degree.
The Laplace transform can be henceforth inverted by computing the roots of such a polynomial, say $\bar{z}_{\rm k}$, with $k=1,...,n+1$, and finally one gets
\begin{equation}
G_2(t)=\sum^{n+1}_{k=1}a_{\rm k}e^{\bar{z}_{\rm k}t},
\end{equation}   
where the $a_{\rm k}$ are constants depending on the roots $\bar{z}_{\rm k}$.
However, one has to notice that some roots have a positive real part, corresponding to divergent run-away solutions~\cite{Ford1991}. 
These divergent roots have to be dropped out. In fact it is possible to see that they do not satisfy the condition $z\ll\Lambda$ that we assumed in the beginning of the calculation. Then,  they are negligible in the long-time limit. 
One may invert the Laplace transform in Eq.~(\ref{LTG2Un}) through the general approach of the Bromwich integral.
Here the run-away roots remain out of the integration contour, and therefore the related residuous do not contribute. 
We conclude that Eq.~(\ref{G2LongTime}) represents the complete expression for the Laplace transform in Eq.~(\ref{LTG2Un}) in the long-time limit. 

\begin{figure}
\begin{center}
\includegraphics[width=0.95\columnwidth]{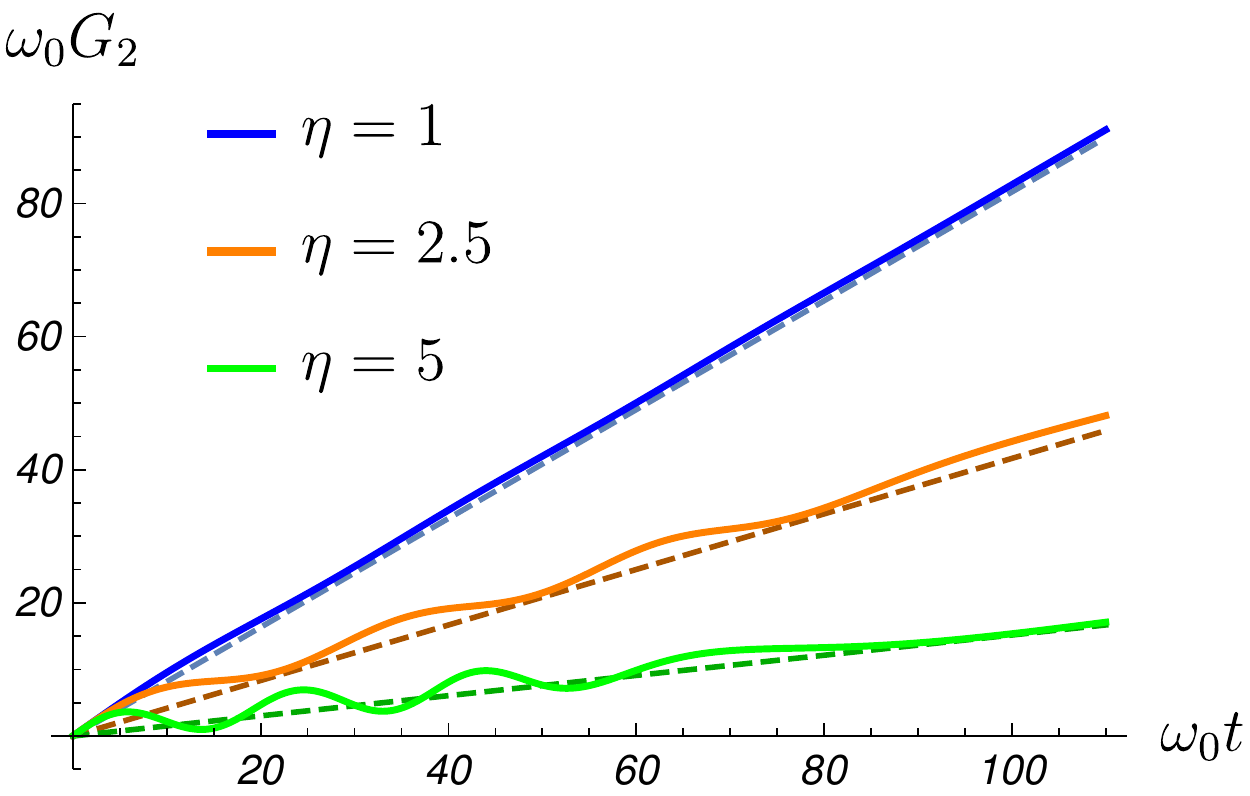}
\caption{\label{G2UntrapNumAn}Inverse Laplace transform of Eq.~(\ref{LTG2Un}). 
The plot refers to a K impurity embedded in a gas of Rb atoms with a density $n_{\rm 0}=7\cdot10^{6}\text{m}^{-1}$ and a coupling constant $g_{\rm B}=2.36\cdot10^{-37}\text{J}\cdot\text{m}$. Here $\omega_0=\hbar n^2_0/m_{\rm I}$ represents the characteristic frequency.
Solid lines refer to  result of the approximation in Eq.~(\ref{Zakian}) for $N=5$, while dashed ones represent analytic long-time predictions from Eq.~(\ref{G2LongTime}). } 
\end{center}
\end{figure}

In order to go beyond any asymptotic expansion of Eq.~(\ref{LTGamma}), we have to perform numerically the inversion of the Laplace transform in Eq.~(\ref{LTG2Un}).
There exist several algorithms aimed to deal with this problem, as carefully discussed in~\cite{Wang2015}. 
Here, we consider the Zakian method, where the inverse Laplace transform $f(t)$ of a function $F(z)$ is approximated as
\begin{equation}\label{Zakian}
\tilde{f}(t)=\frac{2}{t}\sum^{N}_{j=1}\text{Re}\left[k_{\rm j}F\left(\alpha_{\rm j}/t\right)\right],
\end{equation} 
with $\alpha_{\rm j}$ and $k_{\rm j}$  constants that can be either complex or reals. 
In the limit $N\rightarrow\infty$ it turns that $\tilde{f}(t)\rightarrow f(t)$. 
Many studies show that the error in the approximation is negligible in several situations already for $N=5$, and the parameters $\alpha_{\rm j}$ and $k_{\rm j}$ are listed in Table 1 of~\cite{Wang2015}. 
Applying this method to our problem we invert the Laplace transform in Eq.~(\ref{LTG2Un}) without performing any asymptotic approximation. The result is presented in Fig.~\ref{G2UntrapNumAn}.
We note that in the long-time limit the numerical solution matches the analytic one presented in Eq.~(\ref{G2LongTime}).
This linear divergence is approached through damped oscillations, which characterize the short- and middle-time  regimes. 
The same behavior is detected by reproducing the calculation by means of other algorithms in~\cite{Wang2015}, such as the Fourier and Week ones. 

In conclusion, the position operator in the Heisenberg picture for an untrapped impurity is described by the expression in Eq.~(\ref{XHeis}), where $G_1$ is given by Eq.~\eqref{G11}, while $G_2$ is represented in Fig.~\ref{G2UntrapNumAn}. 
Only in the long-time limit it is possible to exhibit an analytic expression for the second function, Eq.~\eqref{G2LongTime}.
Such a term shows a ballistic form, namely it is proportional to time. 
It means that, as time flows the position impurity becomes larger and larger.
The untrapped impurity does not approach the equilibrium, but runs away from its initial position. This  is a reasonable behavior when we remove trapping. 
It is natural to characterize quantitatively the motion of the untrapped impurity with the mean square displacement (MSD)
\begin{equation}
\mbox{MSD}(t)=\ave{\left[x(t)-x(0)\right]^2},
\end{equation}
which provides the deviation between the position at time $t$ and the initial one.  
In experiments dealing with ultracold gases such a quantity can be measured~\cite{Catani2012}.
In the long-time limit it is possible write
\begin{align}\label{d(t)1}
&\mbox{MSD}(t)=\left(\frac{t}{\alpha}\right)^2\ave{ \dot{x}(0)^2}\nonumber\\
+&\frac{1}{2\left(\alpha m_{\rm I}\right)^2}\!\int^t_0\!ds\int^t_0\!d\sigma(t-s)(t-\sigma)\ave{\{B(s),B(\sigma)\}},
\end{align}
where we considered a factorizing initial state $\rho(t)=\rho_S(0)\otimes\rho_B$. The initial conditions of the impurity and  bath oscillators are then uncorrelated. Then,  averages of the form $\ave{\dot{x}(0)B(s)}$ vanish.  
The second term on the right-hand side of Eq.~(\ref{d(t)1}) can be treated noting that:
\begin{equation}
\ave{\{B(s),B(\sigma)\}}=2\hbar\nu(s-\sigma)
\end{equation}
and remembering the definition of the noise kernel in Eq.\ \eqref{NoiseKernel} it turns:
\begin{align}
\nu(s-\sigma)=m_{\rm I}\tilde{\tau}\int^{\Lambda}_0d\omega\cos\left[(s-\sigma)\omega\right]\omega^3\label{NoiseEq},
\end{align} 
where the hyperbolic cotangent in the noise kernel in Eq.~(\ref{NoiseKernel}) has been approximated to one assuming low-temperature. This is an important assumption. 
It is possible to check that for realistic values of the physical quantities such an approximation for the hyperbolic cotangent is reasonable.  
Replacing Eq.~(\ref{NoiseEq}) in the second term of the right-hand side of Eq.~(\ref{d(t)1}), and integrating with respect of time and frequency, it turns
\begin{equation}\label{MSDHT}
\mbox{MSD}(t)=\left[\ave{ \dot{x}(0)^2}\!+\!\frac{\hbar\tilde{\tau}\Lambda^2}{2m_{\rm I}}\right]\left(\frac{t}{\alpha}\right)^2.
\end{equation}
When this quantity shows a linear dependence on time, the impurity experiences normal  diffusion~\cite{BreuerBook}.
Conversely,  the MSD  is proportional to the square of time in Eq.~(\ref{MSDHT}).  
Such a behavior is termed super-diffusion and provides a key signature of the motion of the impurity.

In this context super-diffusion is a consequence of the presence of memory effects.  
In~\cite{BreuerBook} a similar calculation is performed considering an ohmic SD, associated to the absence of memory effects, and leads to a diffusive behavior.
Super-diffusion in Eq.~(\ref{MSDHT}) arises due to the super-ohmic character of the SD. Therefore, it represents a witness of memory effects for a measurable observable. 

Apart from the position, the result we presented above permits to obtain an expression for the momentum of the impurity in the Heisenberg picture. 
In fact, by inserting within Eq.~\eqref{EqX} the expression in Eq.~\eqref{XHeis} with the $G_2$ function in Eq.~\eqref{G2LongTime}, one infers 
\begin{equation}\label{PHeis}
p(t)=\frac{1}{\alpha}\left[p(0)+\int^t_0B(s)ds\right].
\end{equation} 
Equation~\eqref{PHeis} can be used to compute the average energy of an untrapped impurity
\begin{equation}
E(t)=\frac{\ave{p^2(t)}}{2m_{\rm I}}. 
\end{equation}
Proceeding as in the derivation of the MSD we find in the low temperature limit
\begin{align}\label{energyHT}
E^{LT}(t)=&\eta g_{\rm B}n_{\rm 0}+\frac{E(0)}{\alpha^2}+\frac{\hbar\tilde{\tau}}{2\alpha^2}\Lambda^2\nonumber\\
-&\frac{\hbar\tilde{\tau}}{(\alpha t)^2}\left[\cos\left(\Lambda t\right)+\Lambda t\sin\left(\Lambda t\right)-1\right],
\end{align}
where we recover the mean-field energy term, represented now by the first term on the right hand side.
\begin{figure}
\begin{center}
\includegraphics[width=0.95\columnwidth]{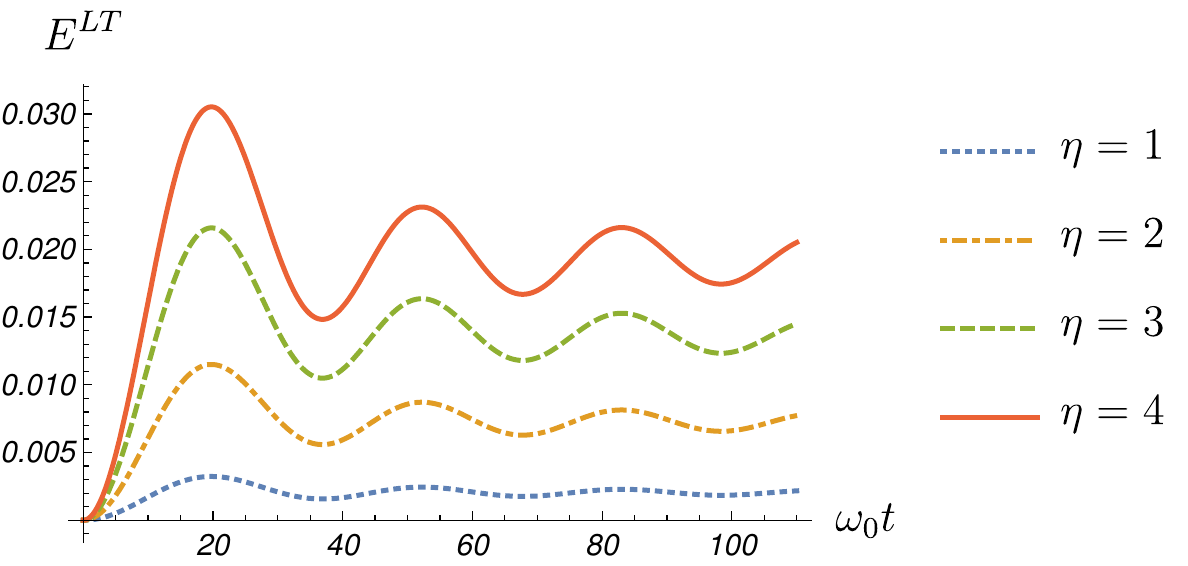}
\caption{\label{energyHTPlot}Average of the energy in the low temperature limit in Eq.~\eqref{energyHT} as a function of the time for different values of the coupling strength. The plot refers to the situation where the initial average energy E(0), as well as the mean-field term, have been set to zero.
}
\end{center}
\end{figure}
Such a quantity is plotted in Fig.~\ref{energyHTPlot}, where it is shown that  the average of the energy oscillates initially and after a long time approaches a constant value. 
These oscillations represent a non-monotonic behavior of the energy. Importantly, the increasing parts correspond to a flow of energy directed from the environment towards the impurity. 
In~\cite{Guarnieri2016} such a backflow energy has been studied for QBM and indicated as a witness of memory effects.  

We note that after a long time, in the weak coupling limit, the only term surviving in Eq.\ \eqref{energyHT} is the mean-field one. In fact the terms in the second line vanish after a long time. The second term on the right hand-side vanish in the weak coupling limit, since it is proportional to $\eta^2$.
In the end, in the weak coupling limit, the asymptotic value of the energy approaches the mean-field one.

In the high-temperature limit  super-diffusion equally emerges, though with a different coefficient. Also the energy shows qualitatively the same behavior as in the low-temperature limit. 
 Nevertheless,  the linear approximation in Eq.~(\ref{mainAssumption}) fails, as discussed  in Appendix~\ref{ValidityApp}.

\section{Trapped Impurity}\label{TrappedSec}
In this section we restore the presence of a harmonic trap, i.e. we look to the situation where $\Omega>0$ in Eq.~(\ref{HarmonicPotential}).
To investigate the dynamics of the impurity we proceed as in the previous section, namely we invert the Laplace transforms in Eqs.~\eqref{LTG1} and~\eqref{LTG2}, in order to characterize the expression of the position operator in the Heisenberg picture, given by in Eq.~\eqref{XHeis}.
Here, an important difference with Sec.~\ref{UntrappedSec} is that the expression for $G_{\rm 1}(t)$ cannot be obtained regardless of any information about the analytic structure of the damping kernel. Conversely, for a trapped impurity the calculation of $G_1$ requires the same attention of $G_2$. 

We approach the problem from a numerical point of view, with the Zakian method presented in~\cite{Wang2015}, briefly described in Sec.~\ref{UntrappedSec}. 
\begin{figure}
\begin{center}
\includegraphics[width=0.95\columnwidth]{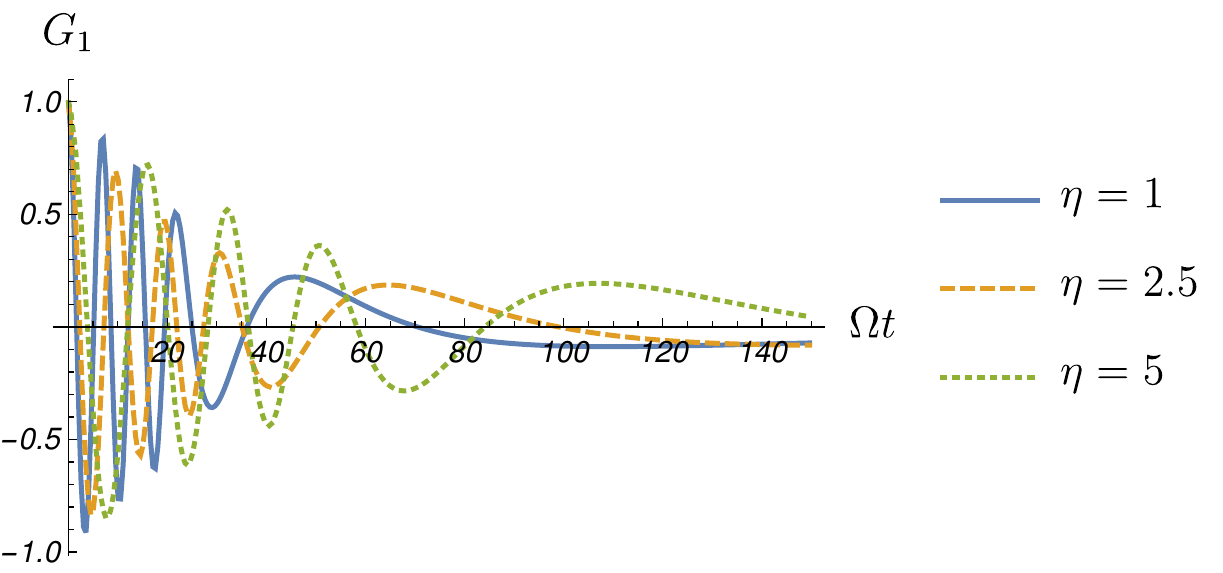}
\includegraphics[width=0.95\columnwidth]{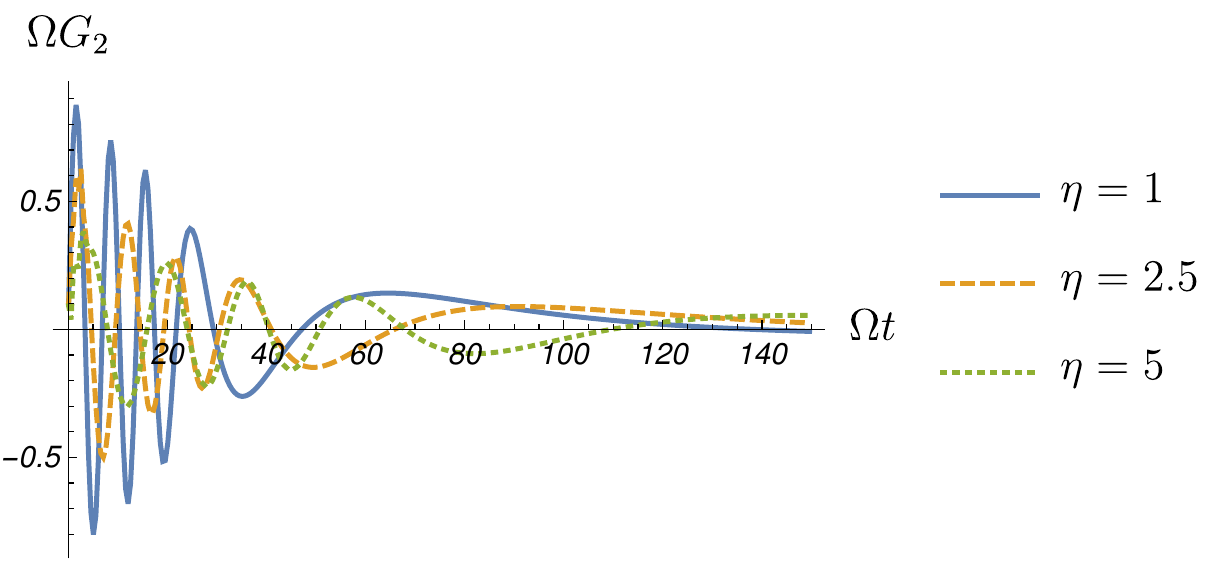}
\caption{\label{GFunctions}Inverse Laplace transforms of Eqs.~\eqref{LTG1} (Up) and~\eqref{LTG2} (Down) obtained through the Zakian method.
The figures refer to an impurity 
of K with $\Omega=2\pi\cdot500$ Hz in a gas made up by Rb with a density of $n_{\rm 0}=7 (\mu\text{m})^{-1}$ and a coupling strength $g_{\rm B}=2.36\cdot10^{-37}$J$\cdot$m.
}
\end{center}
\end{figure}
The results are presented in Fig.~\ref{GFunctions}. 
Both functions exhibit oscillations which get damped in the long-time regime.
In particular, in the long-time limit one gets 
\begin{equation}\label{LongTimeLimitG}
\lim_{t\rightarrow\infty}G_1(t)=0,\quad\lim_{t\rightarrow\infty}G_2(t)=0.
\end{equation}  
The boundary conditions in Eqs.~\eqref{InCondG1} and~\eqref{InCondG2} are satisfied. 
Employing alternative numerical methods described in~\cite{Wang2015} to invert the Laplace transforms we obtain the same result. 
From the physical point of view it means that, after a long-time, the contributions of the initial position and velocity vanish. 

For a trapped impurity we are also interested in the  long-time  behavior of the particle.  
We characterize it by means of the position variance, which is a measurable quantity. 
Taking into account the behavior of $G_1$ and $G_2$ functions showed in Fig.~\ref{GFunctions}, the expression for the position variance can be easily found starting from Eq.~\eqref{XHeis} 
\begin{equation}
\ave{x^2(t)}=\int^t_0ds\int^t_0d\sigma G_2(t-s)G_2(t-\sigma)\frac{\nu(s-\sigma)}{m^2_{\rm I}\hbar^{-1}}, 
\end{equation}
where we used the first line in Eq.~\eqref{NoiseEq}.
Replacing the expression for the noise kernel one gets
\begin{align}
\ave{x^2(t)}&\!=\!\frac{\hbar}{m^2_{\rm I}}\int^{\Lambda}_0 J(\omega)\coth\left(\hbar\omega/2k_{\rm B}T\right)d\omega\nonumber\\
&\!\times\int^t_0ds\int^t_0d\sigma G_2(t-s)G_2(t-\sigma)\cos[\omega(s-\sigma)]\nonumber\\
&\equiv\frac{\hbar}{m^2_{\rm I}}\int^{\Lambda}_0 J(\omega)\coth\left(\hbar\omega/2k_{\rm B}T\right)\phi(t,\omega)d\omega.\label{X2Trap1}
\end{align}
We stress that once the long-time conditions in Eq.~(\ref{LongTimeLimitG}) are fulfilled, it is not necessary to suppose that the state of the global system is initially non-correlated, as we did in Sec.~\ref{UntrappedSec}. 
Equation~\eqref{X2Trap1} turns into
\begin{align}
\phi(t,\omega)=&\frac{1}{2}\int^t_0ds\int^t_0d\sigma G_2(t-s)G_2(t-\sigma)\nonumber\\
\times&\left[e^{i\omega s}e^{-i\omega\sigma}+c.c.\right]\nonumber\\
=&\frac{1}{2}\int^t_0d\tilde{s}  e^{-i\omega\tilde{s}}G_2(\tilde{s})\\
\times&\int^t_0d\tilde{\sigma}e^{i\omega\tilde{\sigma}} G_2(\tilde{\sigma})+c.c.,
\end{align}
where we introduced 
\begin{equation}
\tilde{s}=t-s,\quad\tilde{\sigma}=t-\sigma.
\end{equation}
We are interested in the long-time limit, $t\rightarrow\infty$. 
In this limit, one gets 
\begin{equation}\label{chi2}
\phi(t,\omega)=\mathcal{L}_{-i\omega}\left[G_2(t)\right]\mathcal{L}_{+i\omega}\left[G_2(t)\right]. 
\end{equation}
Replacing Eq.\ \eqref{LTG2} into Eq.\ \eqref{chi2}, we  obtain the final expression for the position variance:
\begin{equation}
\ave{x^2}=\frac{\hbar}{2\pi}\int^{+\Lambda}_{-\Lambda}d\omega\coth\left(\hbar\omega/2k_{\rm B}T\right)\tilde{\chi}''(\omega),\label{X2Trap2}
\end{equation}
where
\begin{equation}\label{ResponseFunction}
\tilde{\chi}''(\omega)=\frac{1}{m_{\rm I}}\frac{\zeta(\omega)\omega}{\left[\omega\zeta(\omega)\right]^2+
\left[\Omega^2-\omega^2+\omega\theta(\omega)\right]^2},
\end{equation}
and
\begin{align}\label{RealAndImLTDamp}
&\zeta\left(\omega\right)=\text{Re}\{\mathcal{L}_{\tilde{z}}\left[\Gamma(t)\right]\}=\frac{\pi\tilde{\tau}}{2}\omega^2+o\left(\frac{\omega}{\Lambda}\right)^5,\\
&\theta\left(\omega\right)=\text{Im}\{\mathcal{L}_{\tilde{z}}\left[\Gamma(t)\right]\}=-\tilde{\tau}\Lambda\omega+\frac{\tilde{\tau}}{\Lambda}\omega^3+o\left(\frac{\omega}{\Lambda}\right)^5. 
\end{align}
with $\tilde{z}=-i\omega+0^{+}$. 
The expression in Eq.~\eqref{X2Trap2}, endowed by Eqs.~\eqref{ResponseFunction} and~\eqref{RealAndImLTDamp}, completely determines the position variance for an impurity trapped in a harmonic potential. 
We emphasize that such an expression has been obtained just by considering the long-time limit of the solution of the Heisenberg equations in Eq.~\eqref{XHeis}. 
It is possible to note, however, that it corresponds to that achieved in the context of the linear response theory by means of the fluctuation-dissipation theorem~\cite{BreuerBook}, as discussed in detail in Appendix \ref{FDRAppendix}.
Indeed, $\tilde{\chi}''$ can be seen as the imaginary part of the Fourier transform of the linear response to an external force applied to the system, at the equilibrium. 

In conclusion, in presence of a harmonic trap the impurity approaches   the equilibrium in the long-time limit. 
We describe such a state through position and momentum variance (a similar expression to Eq.~\eqref{X2Trap2} is also found for the momentum). 
In particular we look to
\begin{equation}\label{DLessVar}
\delta_x=\sqrt{\frac{2m_{\rm I}\Omega\ave{x^2}}{\hbar}},\quad\delta_p=\sqrt{\frac{2\ave{p^2}}{m_{\rm I}\hbar\Omega}},
\end{equation}
which represents the position and momentum variances regularized in order to be dimensionless.  
In these units the Heisenberg principle reads as $\delta_x\delta_p\geq1$, so the Heisenberg threshold is set to be equal to one. 
These quantities do not depend on time, because they refer to an equilibrium stationary state. We study the dependence of $\delta_x$ on the system parameters, such as temperature and interaction strength, that can be tuned in experiments. 

\begin{figure}
\begin{center}
\includegraphics[width=0.95\columnwidth]{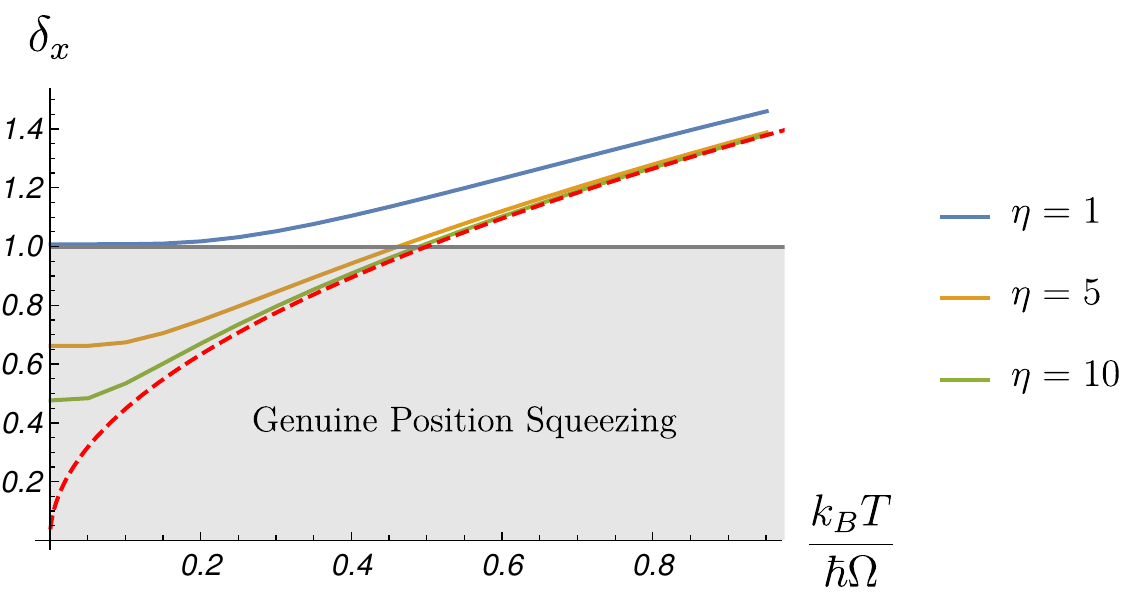}
\caption{\label{posVar}Dimensionless position variance in Eq.~\eqref{DLessVar} as a function of the temperature, for several values of the interaction strength. 
The red dashed line represents the function $\sqrt{2T}$, associated to the equipartition theorem. 
The figure refers to an impurity: 
of K with $\Omega=2\pi\cdot500$ Hz in a gas made up by Rb with a density of $n_{\rm 0}=7 (\mu\text{m})^{-1}$ and a coupling strength $g_{\rm B}=2.36\cdot10^{-37}$J$\cdot$m.
}
\end{center}
\end{figure}

In Fig.~\ref{posVar} we show the position variance for a trapped impurity as a function of the temperature, for several values of the coupling strength.
Such a result follows from both a numerical and analytic integration. 
In the second case, one may proceed by noting that the integrand function rapidly vanishes as $\omega$ increases, so we approximate Eq.~\eqref{X2Trap2} with an integral over the whole real axis (also note that the integrand is an even function). Therefore it is possible to apply the Residue theorem. 
We also stress that Eq.~\eqref{X2Trap2} refers to the asymptotic expansion in Eq.~\eqref{RealAndImLTDamp}, justified in the long-time limit. Such an expansion has been performed till the fifth order in $\omega/\Lambda$, but even going to higher orders we recover the same result as in Fig.~\ref{posVar}.   

In physical grounds we note from  Fig.~\ref{posVar} that for $k_{\rm B}T/\hbar\Omega\gtrsim0.5$, the position variance grows as the square root of the temperature.
Indeed, we detect the behavior provided by $\delta_x\sim\sqrt{2T}$ (red dashed line), in agreement with the equipartition theorem. 
We note that as the temperature increases the value of the position variance turns to be less dependent of the interaction strength. 
 
In the other limit, that is when one approaches the zero-temperature limit  (for $k_{\rm B}T/\hbar\Omega\lesssim0.5$),  we find that $\delta_x<1$. This means that the position variance is smaller than the value related to the Heisenberg threshold in these units. This important effect is named genuine position squeezing, and corresponds to high spatial localization of the impurity.   
\begin{figure}
\begin{center}
\includegraphics[width=0.95\columnwidth]{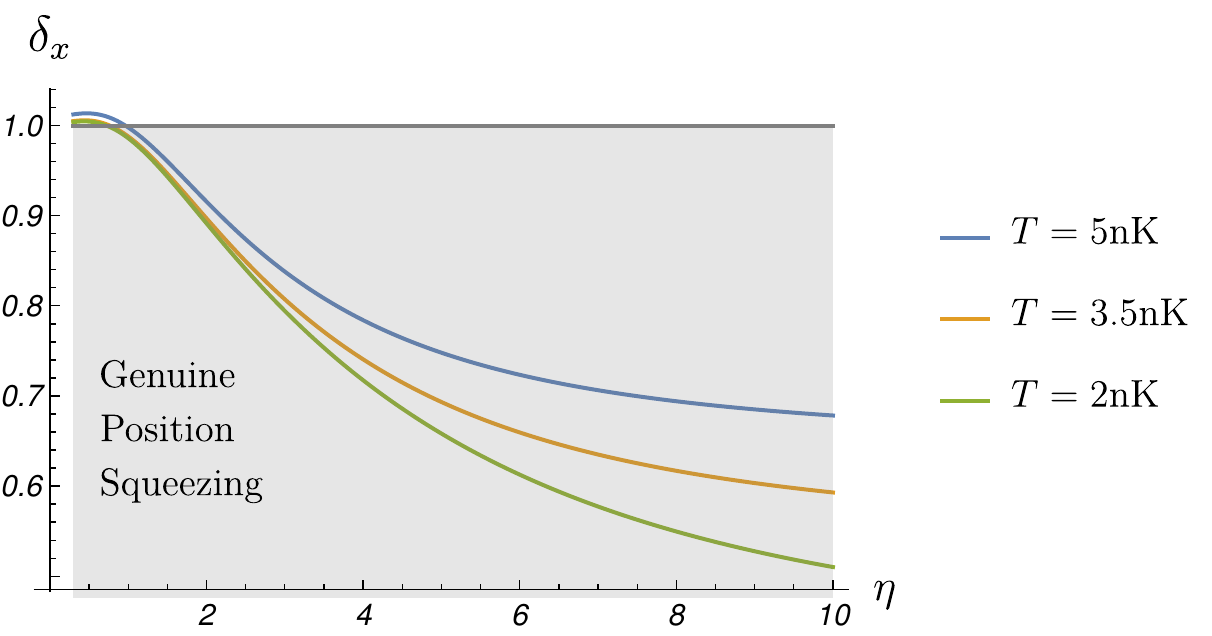}
\caption{\label{minSq}Dimensionless position variance in Eq.~\eqref{DLessVar} as a the interaction strength, for several values of the temperature, satisfying $k_{\rm B}T/\hbar\Omega\ll0.5$. 
The figure refers to an impurity: 
of K with $\Omega=2\pi\cdot500$ Hz in a gas made up by Rb with a density of $n_{\rm 0}=7 (\mu\text{m})^{-1}$ and a coupling strength $g_{\rm B}=2.36\cdot10^{-37}$J$\cdot$m.
}
\end{center}
\end{figure}
This is an important resource in quantum technologies, and therefore we would like to find under which parameters such an effect is enhanced.  In Fig.~\ref{minSq} we plot the value of the dimensionless position variance in Eq.~\eqref{DLessVar} as a function of the interaction strength, obtained by exploring values of the temperature as low as possible. 
We point out that genuine squeezing is maximized, i.e. $\delta_x$ gets smaller, if the temperature tends to zero, and the interaction becomes stronger. For instance, tuning $\eta\approx5$ and setting $T=2$nk it is possible to reach $\delta_x\approx0.7$.
This degree of squeezing may be enhanced by increasing the value of $\eta$. In this case, our results would lie at the border of the validity regime of our Hamiltonian model. 
Note that in experiments the value of $\eta$ may be pushed towards very high values, for instance $\eta=30$ \cite{Catani2012}.
Of course, the present theory cannot be employed to investigate such a regime.  
However, we remark that for $\eta\lesssim7$, where the theory is well defined, we already find a good degree of genuine position squeezing.
We underline, anyway, that even for $\eta\lesssim7$, the investigation of such a limit would be not possible through the tools developed in~\citep{Massignan2005}, basically due to the violations of the Heisenberg principle at low temperature. Here, instead, Heisenberg principle is fulfilled.

Finally,  we would like to underline that the genuine position squeezing appears both for attractive and repulsive interactions. 
This is in agreement with the results presented in \cite{Catani2012} (see Fig.\ 4), where it has been shown that the position variance of the impurity does not depend on the sign of the interaction. 
To understand this, let us first note that  in the  presence of a trap confining the BEC, the density of the condensate, will be spatially changing and  peaked in the center of the trap. If impurity-boson interactions are repulsive,  one should expect that the impurity will be pushed away from the center and localized around the distance $D$ resulting from am interplay between the force trapping the impurity and the force resulting from the mean field  interactions.  While due to the parity symmetry   $\langle x\rangle=0$, one should expect that $ \langle x^2\rangle  \simeq D^2$ and, unless $D$ is very small, there will be no squeezing of the position. In contrast, for attractive interactions the impurity will be localized in the center of the trap and squeezing will be possible.  In Ref.~\cite{Lim2017} it is considered the model case when the bath is coupled to the square of the position of the impurity. This is the first step towards the study of the full problem, in which impurity couples to a confined condensate with spatially dependent density and complicated spatially dependent Bogoliubov-de Gennes modes.  

In contrast, in the present case we consider a spatially homogeneous condensate with constant density and Bogoliubov-de Gennes modes, which are plane waves.
Moreover, we linearize the spatial dependence of modes, assuming self-consistently that the impurity is localized in the region of $x$  allowing for such a linearization. The above arguments do not apply in thus case, and  our results: i)  do not depend on the sign of impurity-bath interaction; ii)  squeezing is thus possible for both repulsive and attractive interactions. 

\section{Conclusions}
We presented a discussion concerning the physics of an impurity embedded in a homogeneous BEC, adopting an open quantum system point of view, in particular recalling the paradigmatic quantum Brownian motion model. 
In this framework the impurity may be treated as a Brownian particle interacting with a bath of Bogoliubov modes. 
Such an approach is suitable to describe the dynamics of the impurity, namely to characterize its motion. 
We pursued this task employing Heisenberg equations, finding an equation for the impurity position (see Eq.~\eqref{EqDiffFin}) which can be seen as the quantum analogue of the stochastic Langevin equation.
Equation~\eqref{EqDiffFin} is non-local in time, suggesting that the dynamics of an impurity in a homogeneous BEC carries a certain amount of memory effects. 
The presence of memory effects is embodied in the structure of the spectral density, which we calculated explicitly in Sec.~\eqref{SDSec}. The spectral density shows a super-ohmic behavior.  

The solution of the equation for the impurity position, together with the spectral density, permits to qualify the motion of the impurity. 
We distinguished two situations: the case where the impurity is trapped in a harmonic potential (Sec.~\ref{TrappedSec}) and that in which there is not any trap (Sec.~\ref{UntrappedSec}). 
When the impurity is untrapped it does not approach the equilibrium, and runs away from the point where was initially localized. 
We characterized quantitatively such a behavior by computing the mean square displacement, providing information about the portion of space explored by the impurity during its random motion. This quantity  can be measured in experiments~\cite{Catani2012}. 
Such a quantity shows a super-diffusive form, namely it is proportional to the square of time. 
The super-diffusive behavior is a consequence of the presence of memory effects.
Therefore, this result constitutes a witness of non-Markovianity on a measurable observable.

In~\cite{Guarnieri2016} the presence of memory effects has also been related to a back-flow of energy, directed from the environment to the impurity. We evaluated the average of energy for an untrapped impurity as a function of the time, and we found non-monotonic behavior: the impurity does not just release energy to the environment (dissipation), but also acquires energy from it. 
It is an open question to understand if such a back-flow of energy may be employed  as a resource for quantum technologies. 

In presence of a trap the impurity approaches the equilibrium after a long-time  evolution. 
It is also interesting to study the position and momentum variance of such a final stable state.  
We study the behavior of both quantities as a function of the temperature and of the interaction strength. 
We find that as the coupling gets stronger, and as the temperature tends to zero, the impurity experiences genuine position squeezing, namely its position variance takes values smaller than the Heisenberg principle. 
This feature corresponds to high localization in space of the impurity. 
Such an effect may be optimized by increasing the strength of the coupling between the impurity and the BEC. 
Here, one has to take into account the constraint imposed on the interaction  in order to ensure the validity of the Fr\"{o}hlich Hamiltonian model, consisting of an upper bound on the quantity $\eta$.
Also within the limitations defined by this condition we find a good degree of genuine position squeezing.

Finally, we would like to highlight that the method we developed in this work can be exported to others ultracold systems. 
For instance, in~\cite{Efimikin2013} it has been proved that the dynamics of a bright soliton in a superfluid in one dimension is described by an equation showing the same form of that in Eq.~\eqref{EqDiffFin}. 
The spectral density for the system is in some circumstances proportional to the third power of the frequency of the Bogoliubov modes, in agreement with the current work.

\acknowledgments
Insightful discussion with Victor Galitski, Jan Wehr, Pietro Massignan, and Nils-Eric Guenther are  gratefully  acknowledged. 

This work has been funded by a scholarship from the Programa M\`{a}sters d'Excel-l\`{e}ncia of the Fundaci\'{o} Catalunya-La Pedrera, ERC Advanced Grant OSYRIS, EU IP SIQS, EU PRO QUIC, 
EU STREP EQuaM (FP7/2007-2013, No. 323714), Fundaci\'o Cellex, the Spanish MINECO (SEVERO OCHOA GRANT SEV-2015-0522,  FOQUS FIS2013-46768, FISICATEAMO FIS2016-79508-P), and the Generalitat de Catalunya (SGR 874 and CERCA/Program).

\appendix

\section{Derivation of the equation for the impurity position}\label{DerHE}

In this Appendix we show in detail the calculation leading to Eq.~(\ref{EqDiffFin}). The starting point is constituted by the Eqs.~(\ref{EqX})-(\ref{Eqbdag}).  
The first step is to derive both sides of Eq.~(\ref{EqX}) and to replace the result in Eq.~(\ref{EqP}), thus obtaining an equation just for the position of the impurity
\begin{equation}\label{EqDiff1}
\ddot{x}{(t)}+\Omega^2x(t)=-i\sum_k\frac{\hbar g_{\rm k}}{m_{\rm I}}\left[b_{\rm k}(t)-b^{\dagger}_{\rm k}(t)\right].
\end{equation}
The time dependence of the Bogoliubov modes can be extracted from Eqs.~(\ref{Eqb}) and~(\ref{Eqbdag}).
They are linear but non-homogeneous first-order differential equations. Therefore their solution is the sum of that of the related homogeneous one and a particular integral. The former can be easily obtained since it is just that of a harmonic oscillator,  
\begin{equation}
b_{\rm k}(t)=b_{\rm k}e^{-i\omega_{\rm k} t}+h^{-}_{\rm k}(t),\quad b^\dagger_{\rm k}(t)=b^{\dagger}_{\rm k}e^{+i\omega_{\rm k}t}+h^{+}_{\rm k}(t).
\end{equation}
The quantities $h^{-}_{\rm k}$ and $h^{+}_{\rm k}$ represent the particular solutions of Eqs.~(\ref{Eqb}) and~(\ref{Eqbdag}), that may be expressed as convolution product of the unknown function $x(t)$ and the Green function
\begin{equation}
h^{\pm}_{\rm k}(t)=\int^{t}_0G^{\pm}_{\rm k}(t-s)x(s)ds.
\end{equation}
Then, the problem of solving the Heisenberg equations for the Bogoliubov modes reduces to that of finding the Green function of Eqs.~(\ref{Eqb}) and~(\ref{Eqbdag}). 
The Green function is defined by the equation
\begin{equation}
\dot{G}^{\pm}_{\rm k}(t)\mp i\omega_{\rm k}{G}^{\pm}_{\rm k}(t)=-g_{\rm k}\delta(t),
\end{equation}
which applying Fourier transform turns into
\begin{equation}\label{FTGreen}
\mathcal{F}_{\tilde{\omega}}\left[G^{\pm}_{\rm k}(t)\right]=-\frac{ig_{\rm k}}{\tilde{\omega}\pm\omega}.
\end{equation}
Thus, the problem of determining the Green function is solved if one performs the inversion of the Fourier transform in Eq.~(\ref{FTGreen}). Namely, one has to calculate the following integral
\begin{equation}\label{GreenIntegral}
G^{\pm}_{\rm k}(t)=-\frac{ig_{\rm k}}{2\pi}\mathcal{P}\int^{+\infty}_{-\infty}\frac{e^{i\tilde{\omega}t}}{\tilde{\omega}\pm \omega_{\rm k}}d\tilde{\omega},
\end{equation}
where we introduced the principal Cauchy part $\mathcal{P}$ because otherwise  the integral is not well-defined in $\tilde{\omega}=\pm\omega_{\rm k}$.
\begin{figure}
\includegraphics[width=0.9\columnwidth]{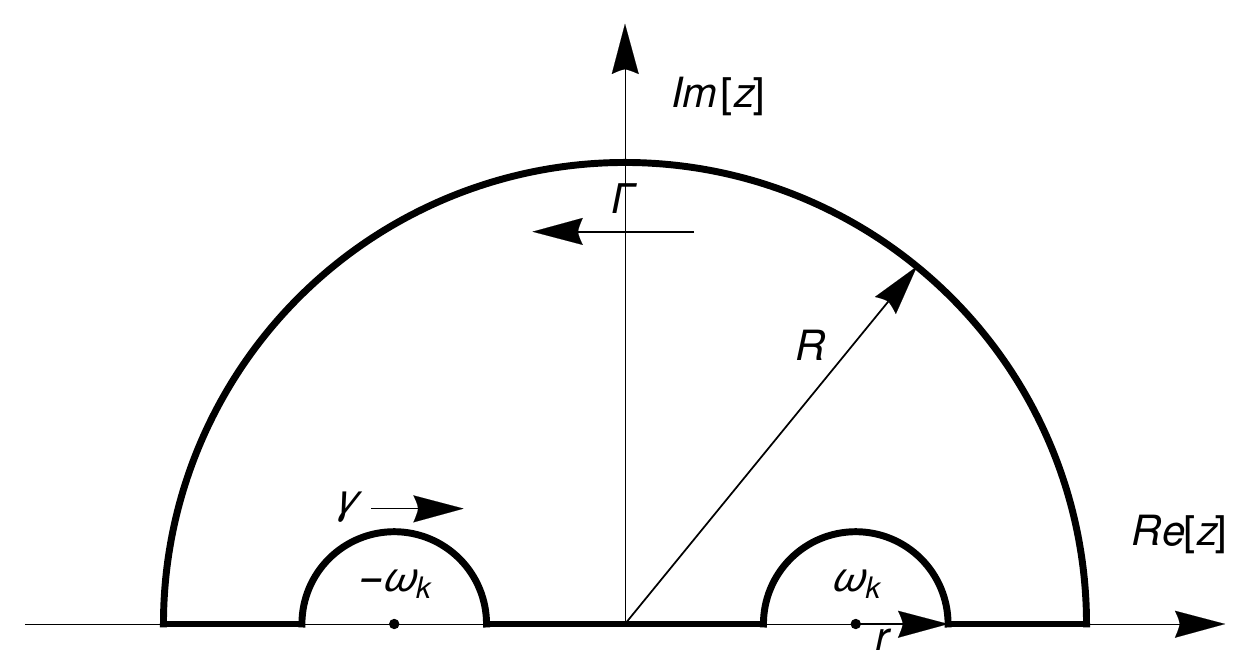}
\caption{\label{PathFigure} Path in complex plane to solve the integral in Eq.\ (\ref{GreenIntegral}).}
\end{figure}
The integral in Eq.~(\ref{GreenIntegral}) can be solved recalling the Jordan Lemma, selecting the path in Fig.~\ref{PathFigure}. 
It follows
\begin{equation}\label{FinalGreen}
G^{\pm}_{\rm k}(t)=\frac{g_{\rm k}}{2}\exp\left(\mp i\omega_{\rm k}t\right).
\end{equation}
In conclusion, Eq.~(\ref{EqDiff1}) takes the form
\begin{align}\label{EqDiff2}
&\ddot{x}(t)+\Omega^2 x(t)-\frac{\hbar}{m_{\rm I}}\sum_kg^2_{\rm k}\int^{t}_0x(s)\sin\left[\omega_{\rm k}(t-s)\right]ds\\
=&\frac{B(t)}{m_{\rm I}},\nonumber
\end{align}
where
\begin{equation}\label{StochasticTerm}
 {B}(t)\equiv-\sum_k\hbar g_{\rm k}{\pi}_{\rm k}(t)=\sum_k i\hbar g_{\rm k}( {b}^{\dagger}_{\rm k}e^{i\omega_{\rm k}t}- {b}_{\rm k}e^{-i\omega_{\rm k}t}). 
\end{equation}
Equation~(\ref{EqDiff2}) can be expressed in terms of the dissipation kernel,  Eq.~(\ref{dampingKernel}) 
\begin{equation}\label{EqDiff3}
\ddot{x}(t)+\Omega^2 x(t)-\frac{1}{m_{\rm I}}\int^{t}_0\lambda(t-s) x(s)ds=\frac{B(t)}{m_{\rm I}}.
\end{equation}
One can also introduce the damping kernel in Eq.~\eqref{DampingKernel}.
The third term in the first hand-side of Eq.~(\ref{EqDiff2}) so writes as
\begin{align}
&-\frac{1}{m_{\rm I}}\int^{t}_0\lambda(t-s) x(s)ds=\int^{t}_0\dot{\Gamma}(t-s)x(s)ds=\\
&=\der{}{t}\int^t_0\Gamma(t-s)x(s)ds-\Gamma(0)x(t)\nonumber.
\end{align}
Accordingly it is possible to express Eq.~(\ref{EqDiff3}) as 
\begin{equation}
\ddot{x}(t)+\tilde{\Omega}^2 x(t)+\der{}{t}\int^{t}_0\Gamma(t-s) x(s)ds=\frac{B(t)}{m_{\rm I}},
\end{equation}
in which we introduced the renormalized frequency
\begin{equation}\label{RenFreq}
\tilde{\Omega}^2=\Omega^2-\Gamma(0).
\end{equation}
Hereafter we neglect the contribution to the frequency provided by $\Gamma(0)$.
This term grows as the interaction strength increase and could lead to a negative renormalized frequency. 
In this context the impurity experiences instability. 
By deriving the equations of motion directly from Hamiltonian 
Eq.~\eqref{HfinLinReno} one avoids the instability problem, at the cost of introducing and additional counter term {\it ad hoc}. With the procedure here presented we can identify in the equations of motion the effect of the absence of such a term.   

\section{Laplace tranform of the damping kernel}\label{DampingKernelAppendix}
In this Appendix we derive the expression of the Laplace transform of the damping kernel presented in Eq.~(\ref{LTGamma}).
We perform the calculation for a cubic SD with a general ultraviolet regularization
\begin{equation}
J(\omega) = m_{\rm I} \tilde{\tau} \omega^3 \Theta(\omega,\Lambda),
\end{equation}
with $\Theta(\omega,\Lambda)>0$ specifying the dependence on the cut-off of the SD. 
The SD in Eq.~(\ref{LTGamma}) corresponds to the particular case when
\begin{equation}\label{sharpCut}
\Theta(\omega,\Lambda) = \theta(\omega-\Lambda),
\end{equation} 
with  $\theta(\cdot)$ the Heaviside step function. 
We will also compute the Laplace transform considering an exponential cut-off 
\begin{equation}\label{expCut}
\Theta(\omega,\Lambda) = \exp\left(-\Lambda\omega\right),
\end{equation} 
showing that a much complicated expression turns out. 

In general, we assume that $\Theta(\omega,\Lambda)$ decays fast enough so that 
\begin{equation}
\label{eq:cutoff1}
\int_0^{\infty} e^{-zt} \left( \int_0^{\infty} |\omega^2\Theta(\omega,\Lambda) \cos(\omega t)| d\omega \right) dt < \infty. 
\end{equation} 
Note that such a behavior covers both the sharp and the exponential dependence on the cut-off. 
Therefore, by Fubini-Tonelli theorem one can interchange the integrals in the following. 
For $z > 0$, it results in
\begin{align}
\mathcal{L}\left[\Gamma\left(t\right)\right]_z &\! =\! \tilde{\tau} \int_0^{\infty} e^{-zt} \left( \int_0^{\infty} \omega^2\Theta(\omega,\Lambda) \cos(\omega t) d\omega \right) dt \\
&=\tilde{\tau} z \int_0^\infty \frac{\omega^2}{\omega^2+z^2}\Theta(\omega,\Lambda) d\omega,
\end{align}
where we used the expression
\begin{equation}
\int^{\infty}_{0}\exp\left(-zt\right)\cos(\omega t)dt=\frac{z}{\omega^2+z^2}, 
\end{equation}
that is the result of an integration by parts. 

Choosing the cut-off function in Eq.~(\ref{sharpCut}), we obtain
\begin{equation}\label{LTdampSharp}
\mathcal{L}\left[\Gamma\left(t\right)\right]_z=z\tilde{\tau}\left[\Lambda-z\arctan\left(\Lambda/z\right)\right],
\end{equation}
as we presented in Eq.\ (\ref{LTGamma}).
If we consider the cut-off function in Eq.~(\ref{expCut}), the Laplace transform of the damping kernel results to be expressed in terms of the Mejer function. Such a function is not suitable to handle in an analytic calculation.
For the environment given by a BEC, the analytical behavior together with the form of the Bogoliubov spectrum justify the  consideration of a sharp cut-off. 

Let us finally note that, regardless of its analytic form, the presence of the cut-off introduced in Eq.\ \eqref{SDcut} does not affect our results.   
This is because  our results refer to the long-time limit, which is not influenced by the high-frequency part of the SD. 
To prove this, one should compare the Laplace transform of the damping kernel induced by the SD in Eq.\ \eqref{SDcut} (shown in Eq.\ \eqref{LTdampSharp}) with that induced by the SD in Eq.\ \eqref{SD1D}, where there is not any cut-off. 
This is enough because such a Laplace transform is the only object carrying information about the SD along all the theory.  

However, the calculation of the Laplace transform of the damping kernel induced by the ``uncutted" SD in Eq.\ \eqref{SD1D} is not an easy task, due to the very complicated form at high-frequency (see Eq.\ \eqref{chi1}). 
It is given by:
\begin{equation}
\mathcal{L}\left[\Gamma\left(t\right)\right]_z=\tilde{\tau} z \int_0^\infty \frac{\omega^2}{\omega^2+z^2}\chi_{1d}(\omega,\Lambda) d\omega.
\end{equation}
We may evaluate it at the first-order in $z$, which is the relevant one at long-times,
\begin{equation}
\mathcal{L}\left[\Gamma\left(t\right)\right]_z=\tilde{\tau} z \int_0^\infty\chi_{1d}(\omega,\Lambda) d\omega+o(z/\Lambda)^2.
\end{equation}
Since
\begin{equation}
\int_0^\infty\chi_{1d}(\omega,\Lambda) d\omega=\Lambda,
\end{equation}
we recover the expression in Eq.\ \eqref{LTdampSharp}, derived with a sharp cut-off. 

In the end, we conclude that at the long-times our results do not depend on whether the SD is cutted or not, i.e. on the existence of the cut-off. 
This could also be inferred by recalling the Tauberian theorem \cite{FloydBook, FellerBook} according to which the long time behavior of a function (in the time domain) is determined by the low frequency behavior of its Laplace transform (in the frequency domain). Since the low frequency behavior of the Laplace transform of the damping kernels for both spectral densities coincides, the long time dynamics of the impurity do not depend on whether there is a cutoff or not.


\section{Validity of the linear approximation}\label{ValidityApp}

In Sec.~\ref{HamSec} we proved that the Hamiltonian of an impurity in a gas may be expressed in the form of that of the QBM. 
A crucial step to perform this task consists of a linear expansion of the exponential appearing in Eq.~\eqref{IntTerm}. 
The present Appendix is devoted to discuss the validity of such an approximation, namely we wonder for which values of the system parameters the condition
\begin{equation}
kx\ll1,\label{condition}
\end{equation}
holds, allowing the expansion in Eq.~\eqref{HIBExp1Ord}. 
Here, $k$ represents the wave number of the Bogoliubov modes, depending on their frequency as showed in Eq.~\eqref{InvBog}. 
This function increases monotonically as
\begin{align}
&k\approx\omega/c\quad\text{for}\quad\omega\ll\Lambda\label{phonons},\\
&k\sim\sqrt{\omega}\quad\text{for}\quad\omega\gg\Lambda\label{quadratic},
\end{align}
hence we can minimize the left-hand side of Eq.~\eqref{condition} by looking to the small frequency regime in Eq.~(\ref{phonons}).
Note that this is in agreement with our treatment, because all the results presented here refer to the phonon linear part of the Bogoliubov dispersion relation. 
In fact such a portion of the Bogoliubov spectrum is embodied in the super-ohmic form of the SD we have considered (see Sec.~\ref{SDSec}).  
Moreover, the ultraviolet sharp cut-off, $\Lambda$, permits to get rid of the contribution due to the non-linear part in Eq.~\eqref{quadratic}.
The condition in Eq.~\eqref{condition} is then
\begin{equation}
\frac{\omega}{c}x\ll1,
\end{equation}
and recalling the expression
\begin{equation}
c=\sqrt{2}\xi\Lambda,
\end{equation}
it turns into
\begin{equation}\label{condition2}
\frac{\omega}{\Lambda}\frac{x}{\xi}\ll\sqrt{2}. 
\end{equation}
We point out that, in order to fulfill the condition, the position of the impurity may acquire large values, provided we consider a frequency much smaller than $\Lambda$. 

The value of the frequency depends in general on the temperature. 
For a system of bosons the energy at a given temperature $T$ is 
\begin{equation}\label{energy}
E(\omega)=\frac{\hbar\omega}{e^{\frac{\hbar\omega}{k_{\rm B}T}}-1}\leq k_{\rm B}T,
\end{equation} 
accordingly one can evaluate Eq.~\eqref{condition2} as 
\begin{equation}\label{condition3}
\frac{k_{\rm B}T}{\hbar c}x\ll1. 
\end{equation}
Note that, because of the inequality in Eq.~\eqref{energy}, the condition in Eq.~\eqref{condition3} provides an upper bound for Eq.~\eqref{condition}. 

We have to evaluate now the position of the impurity. 
The quantity appearing in Eq.~\eqref{condition} is an operator. 
The impurity is modeled as a harmonic oscillator, so its state is Gaussian.
Therefore it ensues
\begin{equation}
x=\ave{x}+\Delta_x=\Delta_x,
\end{equation}
where we considered that, for a harmonic oscillator, the average value of the position is zero. 
Finally, we can state that the linear approximation underlying our analysis is fulfilled if it is satisfied that
\begin{equation}\label{FinalCondition}
\frac{k_{\rm B}T}{\hbar c}\Delta_x\ll1. 
\end{equation}
To evaluate $\Delta_x$, and thus Eq.~\eqref{FinalCondition}, we have to distinguish again the case of trapped and untrapped impurity. 

For the trapped case we consider in Eq.\ (\ref{FinalCondition}) the fluctuation $\Delta_x$ as that given by $\delta_x$ in Eq.~\eqref{DLessVar}, 
\begin{equation}\label{CondTrapEq}
\chi^{(\text{Tr})}\equiv\frac{k_{\rm B}T}{\hbar c}\delta_x(T,\Omega).
\end{equation}
If $\chi^{(\text{Tr})}<1$ it is possible to state that the linear approximation, Eq.~(\ref{mainAssumption}),  is fulfilled.  
\begin{figure}
\begin{center}
\includegraphics[width=0.95\columnwidth]{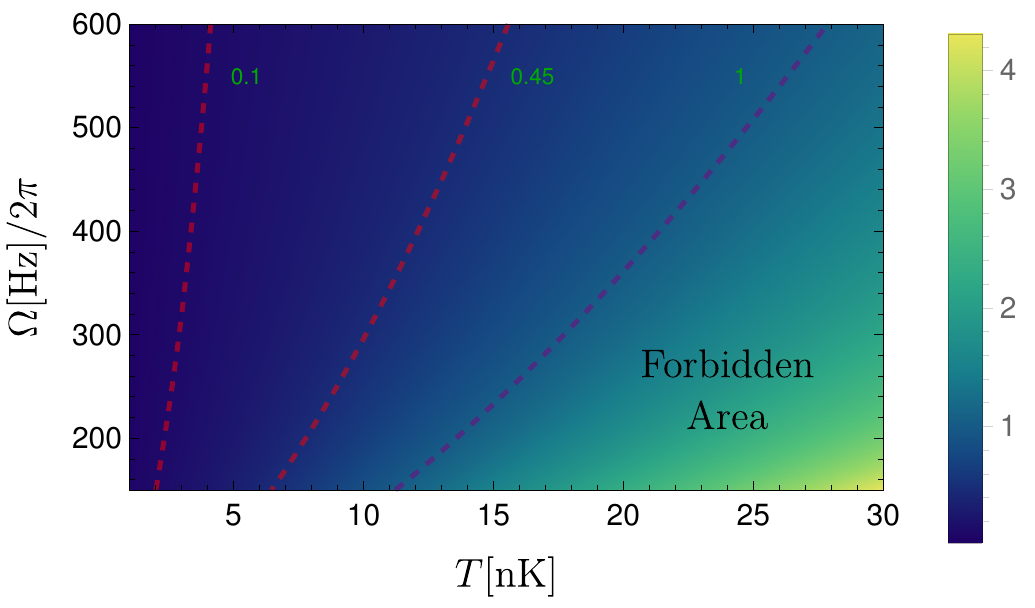}
\caption{\label{condTrap}Behavior of  $\chi^{(\text{Tr})}$ [see Eq.~\eqref{CondTrapEq}] as a function of the temperature and of the frequency of the trap.
 For $\chi^{(\text{Tr})}<1$ the linear approximation, Eq.~(\ref{mainAssumption}), is appropriate for the trapped impurity.  
The figure refers to an impurity 
of K in a gas made up by Rb with a density of $n_{\rm 0}=7 (\mu\text{m})^{-1}$ and a coupling strength $g_{\rm B}=2.36\cdot10^{-37}$J$\cdot$m. The interaction strength is $\eta=1$. }
\end{center}
\end{figure}
We plot in Fig.~\ref{condTrap} the quantity in Eq.~\eqref{CondTrapEq}  as a function of the temperature and the trap frequency.
Note that $\delta_x$ depends also on the interaction strength, but we focus on the case where $\eta=1$.
In general, as the temperature grows the position variance approaches the value predicted by the equipartition theorem, so it gets coupling-independent.  
At low temperature, instead, the value of the position variance approaches one at weak coupling, i.e. $\eta\approx1$, while in general it is smaller (the impurity experiences genuine position squeezing).
Therefore, the value of the position variance at $\eta=1$ represents an upper bound. Therefore,  if the linear approximation is fulfilled for $\eta=1$, it also holds for other, smaller or larger, values of $\eta$. 

Figure~\ref{condTrap} shows that the linear approximation for a trapped impurity is fulfilled at low temperature. 
Moreover, it is very well satisfied as the trap potential gets more and more steep, as pointed out also in~\citep{Cugliandolo2012}.
In particular, we see that for the value we selected to detect genuine position squeezing, i.e. $\Omega=2\pi\cdot500$ and $T\geq2$nm the threshold in Eq.~\eqref{CondTrapEq} takes a value smaller than $0.1$, suggesting that the condition allowing the linear approximation is very well satisfied. 

Next we consider the untrapped impurity. 
In this case $\Delta_x$ may be evaluated by means of the MSD in Eq.~\eqref{MSDHT}. 
Considering the case in which $\ave{\dot{x}^2(0)}=0$, the left-hand side in Eq.~\eqref{condition3} is
\begin{equation}\label{condUntr}
\chi^{(\text{Un})}\equiv\frac{k_{\rm B}T}{\hbar c}\sqrt{\frac{\hbar\tilde{\tau}}{2m_{\rm I}}}\frac{t\Lambda}{\alpha(\eta)}.
\end{equation}  
Again, in order to state that our linear approximation is satisfied the quantity in Eq.~\eqref{condUntr} has to be smaller than one.
\begin{figure}
\begin{center}
\includegraphics[width=0.95\columnwidth]{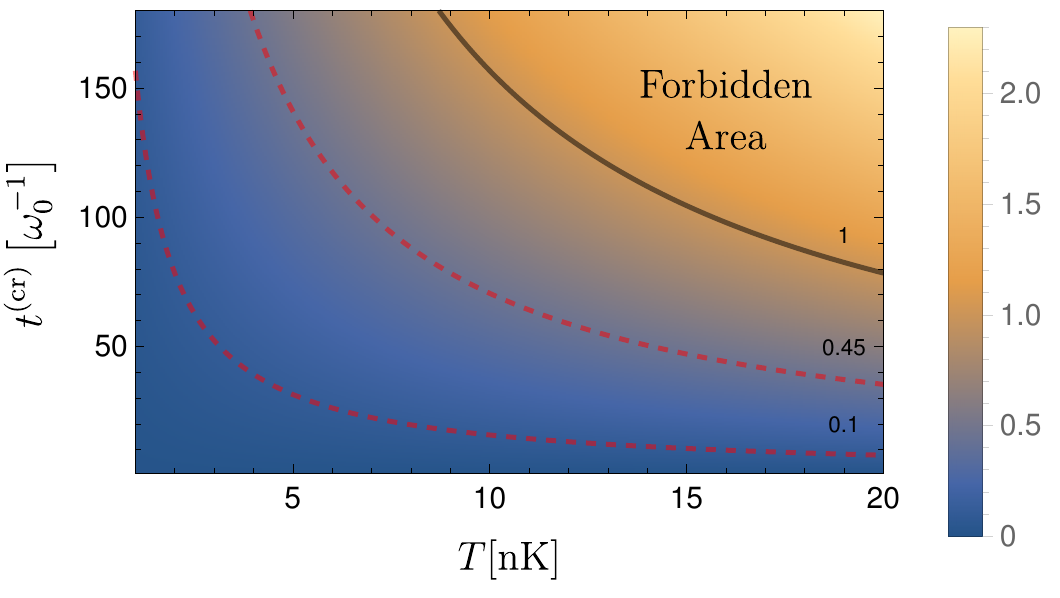}
\caption{\label{condTrapUnd}Behavior of  $\chi^{(\text{Un})}$ [see Eq.~\eqref{condUntr}] as a function of the temperature and of the time.  
For $\chi^{(\text{Un})}<1$ the linear approximation, Eq.~(\ref{mainAssumption}),  is appropriate for the untrapped case. 
The figure refers to an impurity 
of K in a gas made up by Rb with a density of $n_{\rm 0}=7 (\mu\text{m})^{-1}$ and a coupling strength $g_{\rm B}=2.36\cdot10^{-37}$J$\cdot$m. The interaction strength is $\eta=1$. 
}
\end{center}
\end{figure}

In Fig.\ \ref{condTrapUnd} we plotted the value of the quantity $\chi^{(\text{Un})}$ introduced in Eq.\ \eqref{condUntr}.
Even in this case we find that the linear approximation underlying our treatment is very well fulfilled at low-temperature.
For an untrapped impurity the validity of such an approximation holds until a certain value of the time
\begin{equation}\label{tcrit}
t^{(\text{cr})}=\frac{\hbar c}{k_{\rm B}T}\sqrt{\frac{2m_{\rm I}}{\hbar\tilde{\tau}}}\frac{\alpha(\eta)}{\Lambda},
\end{equation}
corresponding to the instant in which the quantity in Eq.~\eqref{condTrap} is equal to one, i.e. we have
\begin{equation}
\chi^{(\text{Un})}(t^{(\text{cr})})=1.
\end{equation}
\begin{figure}
\begin{center}
\includegraphics[width=0.95\columnwidth]{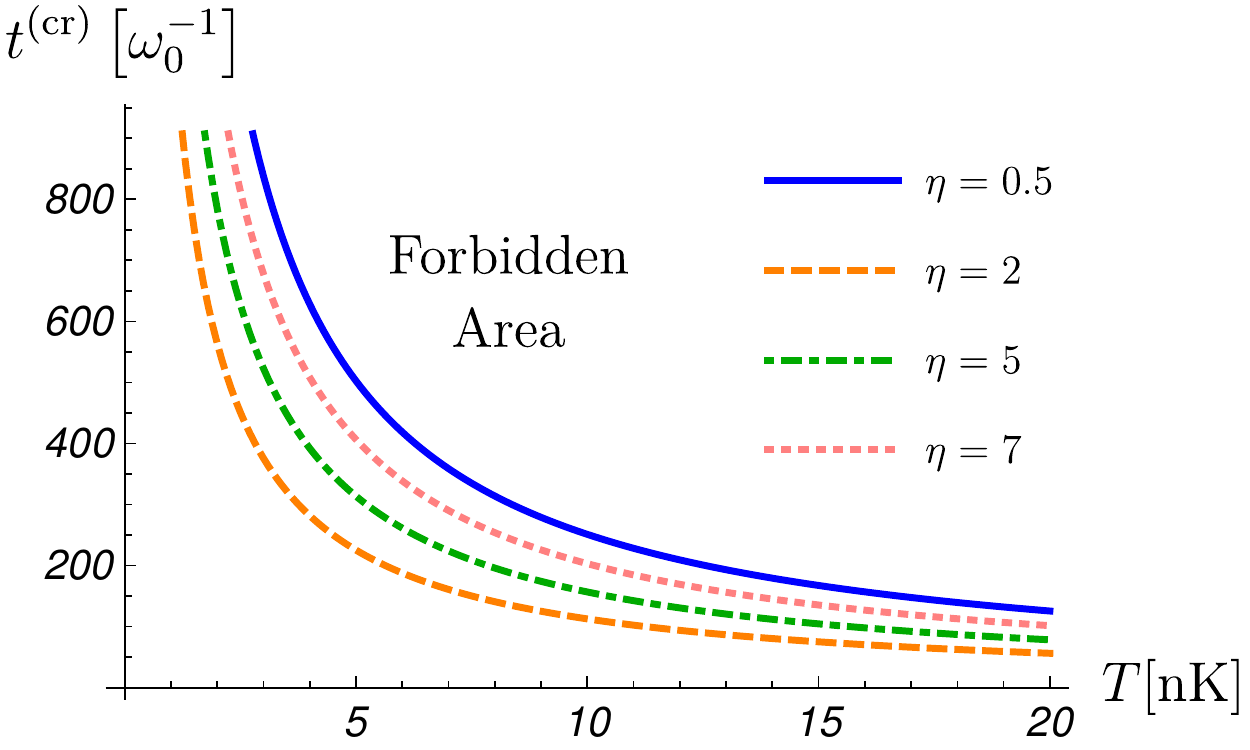}
\caption{\label{critTimePlot}Critical time in Eq.~(\ref{tcrit}) as a function of the temperature for different values of the interaction strength. Below the curves the linear approximation, Eq.~(\ref{mainAssumption}),  is fulfilled. 
The figure refers to an impurity 
of K in a gas made up by Rb with a density of $n_{\rm 0}=7 (\mu\text{m})^{-1}$ and a coupling strength $g_{\rm B}=2.36\cdot10^{-37}$J$\cdot$m.
}
\end{center}
\end{figure}

The critical time in Eq.~\eqref{tcrit} is plotted in Fig.~\ref{critTimePlot} as a function of the temperature, for different values of the interaction strength. 
Each line corresponds to a different value of $\eta$. 
Once one fixes $\eta$, the corresponding line determines the value of time and temperature for which the threshold in Eq.~\eqref{condUntr} is equal to one. 
Accordingly, above the line we have a forbidden area. 
Again, we find that this forbidden area is enlarged as the temperature grows, suggesting that our approximation works at low temperature. 
\section{Derivation of the position variance by means of the fluctuation-dissipation theorem} \label{FDRAppendix}



In this Appendix we present an alternative way to derive the position variance for a trapped impurity, 
employing the linear response theory. 
For this goal we introduce the response function, $\chi(t)=\frac{i}{\hbar} \theta(t)\langle [x(t), x(0)] \rangle$, describing the linear response of the system to an external force at the equilibrium. 
Here $\theta(t)$ is the step function, specifying causality. 
The response function defined here is also known as generalized susceptibility. 
Taking the commutator with $x(0)$ on both sides of \eqref{EqDiffFin} and then averaging (note the initial slip term and the stochastic forcing term are absent once a commutator of $x(0)$ is applied to them), we obtain the following c-valued equation for $\chi(t)$:
\begin{equation}
\ddot{\chi}(t) + \Omega^2 \chi(t) + \int_0^t \Gamma(t-s) \dot{\chi}(s) ds = 0
\end{equation} 
with $\chi(0)=0$, $\dot{\chi}(0) =1/m_{\rm I}$.

Next, consider the symmetric position autocorrelation function, defined as 
\begin{equation}
S(t) = \frac{1}{2}\langle \{x(t), x(0)\} \rangle.
\end{equation}
We recall the fluctuation-dissipation relation of Callen-Welton \cite{BreuerBook}, which relates the equilibrium fluctuation to the response function in the frequency domain as follows:
\begin{equation}
\tilde{S}(\omega) = \hbar \coth \left(\frac{\hbar\omega}{2k_{\rm B}T} \right) \tilde{\chi}''(\omega), 
\end{equation}
where
\begin{equation}
\tilde{S}(\omega) =  \int^{\infty}_{-\infty} dt \cos(\omega t) S(t),
\end{equation}
and
\begin{equation} 
\tilde{\chi}''(\omega) = \text{Im}\{\mathcal{F}_\omega\left[\chi(t)\right] \}
\end{equation}

Using this, we can obtain the symmetric position autocorrelation function:
\begin{equation}
S(t) = \int^{\infty}_{-\infty} d\omega \frac{\hbar}{2\pi} \coth \left(\frac{\hbar \omega}{2k_{\rm B}T} \right) \tilde{\chi}''(\omega) \cos(\omega t).
\end{equation} 
from which the position variance at the equilibrium follows:
\begin{equation}\label{x2FDT}
\ave{x^2}=S(0)= \int^{\infty}_{-\infty} d\omega \frac{\hbar}{2\pi} \coth \left(\frac{\hbar \omega}{2k_{\rm B}T} \right) \tilde{\chi}''(\omega) 
\end{equation}
Note that $\tilde{\chi}''(w)  = \text{Im}\{\mathcal{L}_{\bar{z}}[\chi(t)]\}$, in which 
\begin{align} 
\mathcal{L}_{\bar{z}}[\chi(t)] &= \frac{1}{m_{\rm I}}\mathcal{L}_{\bar{z}}[G_{2}(t)] \\ 
&= \frac{1}{m_{\rm I}} \frac{1}{(\Omega^2 - \omega^2) - i\omega \mathcal{L}_{\bar{z}}[\Gamma(t)]},
\end{align}
where we used the expression in Eq.\ \eqref{LTG2} for the Laplace transform of $G_2$, and $\bar{z}=-\omega+0^+$ as defined in Sec. \ref{TrappedSec}. 
Therefore, we have
\begin{equation}
\tilde{\chi}''(\omega)= \frac{1}{m_{\rm I}} \frac{\omega \xi(\omega) }{\left[\Omega^2-\omega^2-\omega \theta(\omega)\right]^2 + \left[\omega \xi(\omega)\right]^2}.
\end{equation}
This expression has the same form as Eq.~\eqref{ResponseFunction}.
Accordingly, the position variance in Eq.\ \eqref{x2FDT}, obtained recalling the fluctuation-dissipation theorem, corresponds to that in Eq.\ \eqref{X2Trap2}, calculated by solving Heisenberg equation and adopting the SD in Eq.\ \eqref{SDcut}.
Therefore, we prove that the two methods lead to the same result. 

One can proceed in a similar way to evaluate the MSD for an untrapped impurity in the context of linear response theory, rather than solving Heisenberg equation. This method leads however to a very complicated expression for the MSD, that does not provide a more convenient alternative to Eq.\eqref{d(t)1}. This topic is discussed in detail in \cite{Grabert1985}.

\bibliographystyle{plainnat}
\bibliography{QBM_in_a_BEC.bib}

\begin{thebibliography}{65}%
\makeatletter
\providecommand \@ifxundefined [1]{%
 \@ifx{#1\undefined}
}%
\providecommand \@ifnum [1]{%
 \ifnum #1\expandafter \@firstoftwo
 \else \expandafter \@secondoftwo
 \fi
}%
\providecommand \@ifx [1]{%
 \ifx #1\expandafter \@firstoftwo
 \else \expandafter \@secondoftwo
 \fi
}%
\providecommand \natexlab [1]{#1}%
\providecommand \enquote  [1]{``#1''}%
\providecommand \bibnamefont  [1]{#1}%
\providecommand \bibfnamefont [1]{#1}%
\providecommand \citenamefont [1]{#1}%
\providecommand \href@noop [0]{\@secondoftwo}%
\providecommand \href [0]{\begingroup \@sanitize@url \@href}%
\providecommand \@href[1]{\@@startlink{#1}\@@href}%
\providecommand \@@href[1]{\endgroup#1\@@endlink}%
\providecommand \@sanitize@url [0]{\catcode `\\12\catcode `\$12\catcode
  `\&12\catcode `\#12\catcode `\^12\catcode `\_12\catcode `\%12\relax}%
\providecommand \@@startlink[1]{}%
\providecommand \@@endlink[0]{}%
\providecommand \url  [0]{\begingroup\@sanitize@url \@url }%
\providecommand \@url [1]{\endgroup\@href {#1}{\urlprefix }}%
\providecommand \urlprefix  [0]{URL }%
\providecommand \Eprint [0]{\href }%
\providecommand \doibase [0]{http://dx.doi.org/}%
\providecommand \selectlanguage [0]{\@gobble}%
\providecommand \bibinfo  [0]{\@secondoftwo}%
\providecommand \bibfield  [0]{\@secondoftwo}%
\providecommand \translation [1]{[#1]}%
\providecommand \BibitemOpen [0]{}%
\providecommand \bibitemStop [0]{}%
\providecommand \bibitemNoStop [0]{.\EOS\space}%
\providecommand \EOS [0]{\spacefactor3000\relax}%
\providecommand \BibitemShut  [1]{\csname bibitem#1\endcsname}%
\let\auto@bib@innerbib\@empty
\bibitem [{\citenamefont {Landau}\ and\ \citenamefont
  {Pekar}(1948)}]{Landau48}%
  \BibitemOpen
  \bibfield  {author} {\bibinfo {author} {\bibfnamefont {L.~D.}\ \bibnamefont
  {Landau}}\ and\ \bibinfo {author} {\bibfnamefont {S.~I.}\ \bibnamefont
  {Pekar}},\ }\bibfield  {title} {\bibinfo {title} {\emph {Effective mass of a
  polaron}},\ }\href
  {http://www.ujp.bitp.kiev.ua/files/journals/53/si/53SI15p.pdf} {\bibfield
  {journal} {\bibinfo  {journal} {Zh. Eksp. Teor. Fiz.}\ } (\bibinfo {year}
  {1948})}\BibitemShut {NoStop}%
\bibitem [{\citenamefont {Fr{\"o}hlich}(1954)}]{Frolich1954}%
  \BibitemOpen
  \bibfield  {author} {\bibinfo {author} {\bibfnamefont {H.}~\bibnamefont
  {Fr{\"o}hlich}},\ }\bibfield  {title} {\bibinfo {title} {\emph {Electrons in
  lattice fields}},\ }\href
  {http://www.tandfonline.com/doi/abs/10.1080/00018735400101213} {\bibfield
  {journal} {\bibinfo  {journal} {Advances In Physics}\ }\textbf {\bibinfo
  {volume} {3(11):325}} (\bibinfo {year} {1954})}\BibitemShut {NoStop}%
\bibitem [{\citenamefont {Alexandrov}\ and\ \citenamefont
  {Devreese}(2009)}]{Alexandrov2009}%
  \BibitemOpen
  \bibfield  {author} {\bibinfo {author} {\bibfnamefont {A.}~\bibnamefont
  {Alexandrov}}\ and\ \bibinfo {author} {\bibfnamefont {J.}~\bibnamefont
  {Devreese}},\ }\href {http://books.google.es/books?id=EI0Hql-9oY8C} {\emph
  {\bibinfo {title} {Advances in Polaron Physics}}},\ Springer Series in
  Solid-State Sciences\ (\bibinfo  {publisher} {Springer},\ \bibinfo {year}
  {2009})\BibitemShut {NoStop}%
\bibitem [{\citenamefont {Schirotzek}\ \emph {et~al.}(2009)\citenamefont
  {Schirotzek}, \citenamefont {Wu}, \citenamefont {Sommer},\ and\ \citenamefont
  {Zwierlein}}]{Schirotzek2009}%
  \BibitemOpen
  \bibfield  {author} {\bibinfo {author} {\bibfnamefont {A.}~\bibnamefont
  {Schirotzek}}, \bibinfo {author} {\bibfnamefont {C.-H.}\ \bibnamefont {Wu}},
  \bibinfo {author} {\bibfnamefont {A.}~\bibnamefont {Sommer}}, \ and\ \bibinfo
  {author} {\bibfnamefont {M.~W.}\ \bibnamefont {Zwierlein}},\ }\bibfield
  {title} {\bibinfo {title} {\emph {Observation of Fermi Polarons in a Tunable
  Fermi Liquid of Ultracold Atoms}},\ }\href {\doibase
  10.1103/PhysRevLett.102.230402} {\bibfield  {journal} {\bibinfo  {journal}
  {Phys. Rev. Lett.}\ }\textbf {\bibinfo {volume} {102}},\ \bibinfo {pages}
  {230402} (\bibinfo {year} {2009})}\BibitemShut {NoStop}%
\bibitem [{\citenamefont {Kohstall}\ \emph {et~al.}(2012)\citenamefont
  {Kohstall}, \citenamefont {Zaccanti}, \citenamefont {Jag}, \citenamefont
  {Trenkwalder}, \citenamefont {Massignan}, \citenamefont {Bruun},
  \citenamefont {Schreck},\ and\ \citenamefont {Grimm}}]{Kohstall2012}%
  \BibitemOpen
  \bibfield  {author} {\bibinfo {author} {\bibfnamefont {C.}~\bibnamefont
  {Kohstall}}, \bibinfo {author} {\bibfnamefont {M.}~\bibnamefont {Zaccanti}},
  \bibinfo {author} {\bibfnamefont {M.}~\bibnamefont {Jag}}, \bibinfo {author}
  {\bibfnamefont {A.}~\bibnamefont {Trenkwalder}}, \bibinfo {author}
  {\bibfnamefont {P.}~\bibnamefont {Massignan}}, \bibinfo {author}
  {\bibfnamefont {G.~M.}\ \bibnamefont {Bruun}}, \bibinfo {author}
  {\bibfnamefont {F.}~\bibnamefont {Schreck}}, \ and\ \bibinfo {author}
  {\bibfnamefont {R.}~\bibnamefont {Grimm}},\ }\bibfield  {title} {\bibinfo
  {title} {\emph {Metastability and coherence of repulsive polarons in a
  strongly interacting Fermi mixture}},\ }\href
  {http://dx.doi.org/10.1038/nature11065} {\bibfield  {journal} {\bibinfo
  {journal} {Nature}\ }\textbf {\bibinfo {volume} {485}},\ \bibinfo {pages}
  {615} (\bibinfo {year} {2012})}\BibitemShut {NoStop}%
\bibitem [{\citenamefont {{Koschorreck}}\ \emph {et~al.}(2012)\citenamefont
  {{Koschorreck}}, \citenamefont {{Pertot}}, \citenamefont {{Vogt}},
  \citenamefont {{Fr{\"o}hlich}}, \citenamefont {{Feld}},\ and\ \citenamefont
  {{K{\"o}hl}}}]{Koschorreck2012}%
  \BibitemOpen
  \bibfield  {author} {\bibinfo {author} {\bibfnamefont {M.}~\bibnamefont
  {{Koschorreck}}}, \bibinfo {author} {\bibfnamefont {D.}~\bibnamefont
  {{Pertot}}}, \bibinfo {author} {\bibfnamefont {E.}~\bibnamefont {{Vogt}}},
  \bibinfo {author} {\bibfnamefont {B.}~\bibnamefont {{Fr{\"o}hlich}}},
  \bibinfo {author} {\bibfnamefont {M.}~\bibnamefont {{Feld}}}, \ and\ \bibinfo
  {author} {\bibfnamefont {M.}~\bibnamefont {{K{\"o}hl}}},\ }\bibfield  {title}
  {\bibinfo {title} {\emph {{Attractive and repulsive Fermi polarons in two
  dimensions}}},\ }\href {\doibase 10.1038/nature11151} {\bibfield  {journal}
  {\bibinfo  {journal} {Nature}\ }\textbf {\bibinfo {volume} {485}},\ \bibinfo
  {pages} {619} (\bibinfo {year} {2012})}\BibitemShut {NoStop}%
\bibitem [{\citenamefont {Massignan}\ \emph {et~al.}(2014)\citenamefont
  {Massignan}, \citenamefont {Zaccanti},\ and\ \citenamefont
  {Bruun}}]{MassignanPolRev2014}%
  \BibitemOpen
  \bibfield  {author} {\bibinfo {author} {\bibfnamefont {P.}~\bibnamefont
  {Massignan}}, \bibinfo {author} {\bibfnamefont {M.}~\bibnamefont {Zaccanti}},
  \ and\ \bibinfo {author} {\bibfnamefont {G.~M.}\ \bibnamefont {Bruun}},\
  }\bibfield  {title} {\bibinfo {title} {\emph {Polarons, dressed molecules and
  itinerant ferromagnetism in ultracold Fermi gases}},\ }\href
  {http://stacks.iop.org/0034-4885/77/i=3/a=034401} {\bibfield  {journal}
  {\bibinfo  {journal} {Reports on Progress in Physics}\ }\textbf {\bibinfo
  {volume} {77}},\ \bibinfo {pages} {034401} (\bibinfo {year}
  {2014})}\BibitemShut {NoStop}%
\bibitem [{\citenamefont {{Lan}}\ and\ \citenamefont {{Lobo}}(2014)}]{Lan2014}%
  \BibitemOpen
  \bibfield  {author} {\bibinfo {author} {\bibfnamefont {Z.}~\bibnamefont
  {{Lan}}}\ and\ \bibinfo {author} {\bibfnamefont {C.}~\bibnamefont {{Lobo}}},\
  }\bibfield  {title} {\bibinfo {title} {\emph {{A single impurity in an ideal
  atomic Fermi gas: current understanding and some open problems}}},\ }\href
  {http://journal.library.iisc.ernet.in/index.php/iisc/search/advancedResults}
  {\bibfield  {journal} {\bibinfo  {journal} {J. Indian I. Sci.}\ }\textbf
  {\bibinfo {volume} {94}},\ \bibinfo {pages} {179} (\bibinfo {year}
  {2014})}\BibitemShut {NoStop}%
\bibitem [{\citenamefont {{Levinsen}}\ and\ \citenamefont
  {{Parish}}(2014)}]{Levinsen2014}%
  \BibitemOpen
  \bibfield  {author} {\bibinfo {author} {\bibfnamefont {J.}~\bibnamefont
  {{Levinsen}}}\ and\ \bibinfo {author} {\bibfnamefont {M.~M.}\ \bibnamefont
  {{Parish}}},\ }\bibfield  {title} {\bibinfo {title} {\emph {{Strongly
  interacting two-dimensional Fermi gases}}},\ }\href@noop {} {\  (\bibinfo
  {year} {2014})},\ \Eprint {http://arxiv.org/abs/1408.2737} {arXiv:1408.2737}
  \BibitemShut {NoStop}%
\bibitem [{\citenamefont {Schmidt}\ \emph {et~al.}(2012)\citenamefont
  {Schmidt}, \citenamefont {Enss}, \citenamefont {Pietil\"a},\ and\
  \citenamefont {Demler}}]{Schmidt2012}%
  \BibitemOpen
  \bibfield  {author} {\bibinfo {author} {\bibfnamefont {R.}~\bibnamefont
  {Schmidt}}, \bibinfo {author} {\bibfnamefont {T.}~\bibnamefont {Enss}},
  \bibinfo {author} {\bibfnamefont {V.}~\bibnamefont {Pietil\"a}}, \ and\
  \bibinfo {author} {\bibfnamefont {E.}~\bibnamefont {Demler}},\ }\bibfield
  {title} {\bibinfo {title} {\emph {Fermi polarons in two dimensions}},\ }\href
  {\doibase 10.1103/PhysRevA.85.021602} {\bibfield  {journal} {\bibinfo
  {journal} {Phys. Rev. A}\ }\textbf {\bibinfo {volume} {85}},\ \bibinfo
  {pages} {021602} (\bibinfo {year} {2012})}\BibitemShut {NoStop}%
\bibitem [{\citenamefont {C\^ot\'e}\ \emph {et~al.}(2002)\citenamefont
  {C\^ot\'e}, \citenamefont {Kharchenko},\ and\ \citenamefont
  {Lukin}}]{Cote2002}%
  \BibitemOpen
  \bibfield  {author} {\bibinfo {author} {\bibfnamefont {R.}~\bibnamefont
  {C\^ot\'e}}, \bibinfo {author} {\bibfnamefont {V.}~\bibnamefont
  {Kharchenko}}, \ and\ \bibinfo {author} {\bibfnamefont {M.~D.}\ \bibnamefont
  {Lukin}},\ }\bibfield  {title} {\bibinfo {title} {\emph {Mesoscopic Molecular
  Ions in Bose-Einstein Condensates}},\ }\href {\doibase
  10.1103/PhysRevLett.89.093001} {\bibfield  {journal} {\bibinfo  {journal}
  {Phys. Rev. Lett.}\ }\textbf {\bibinfo {volume} {89}},\ \bibinfo {pages}
  {093001} (\bibinfo {year} {2002})}\BibitemShut {NoStop}%
\bibitem [{\citenamefont {Massignan}\ \emph {et~al.}(2005)\citenamefont
  {Massignan}, \citenamefont {Pethick},\ and\ \citenamefont
  {Smith}}]{Massignan2005}%
  \BibitemOpen
  \bibfield  {author} {\bibinfo {author} {\bibfnamefont {P.}~\bibnamefont
  {Massignan}}, \bibinfo {author} {\bibfnamefont {C.~J.}\ \bibnamefont
  {Pethick}}, \ and\ \bibinfo {author} {\bibfnamefont {H.}~\bibnamefont
  {Smith}},\ }\bibfield  {title} {\bibinfo {title} {\emph {Static properties of
  positive ions in atomic Bose-Einstein condensates}},\ }\href {\doibase
  10.1103/PhysRevA.71.023606} {\bibfield  {journal} {\bibinfo  {journal} {Phys.
  Rev. A}\ }\textbf {\bibinfo {volume} {71}},\ \bibinfo {pages} {023606}
  (\bibinfo {year} {2005})}\BibitemShut {NoStop}%
\bibitem [{\citenamefont {Cucchietti}\ and\ \citenamefont
  {Timmermans}(2006)}]{Cucchietti2006}%
  \BibitemOpen
  \bibfield  {author} {\bibinfo {author} {\bibfnamefont {F.~M.}\ \bibnamefont
  {Cucchietti}}\ and\ \bibinfo {author} {\bibfnamefont {E.}~\bibnamefont
  {Timmermans}},\ }\bibfield  {title} {\bibinfo {title} {\emph {Strong-Coupling
  Polarons in Dilute Gas Bose-Einstein Condensates}},\ }\href {\doibase
  10.1103/PhysRevLett.96.210401} {\bibfield  {journal} {\bibinfo  {journal}
  {Phys. Rev. Lett.}\ }\textbf {\bibinfo {volume} {96}},\ \bibinfo {pages}
  {210401} (\bibinfo {year} {2006})}\BibitemShut {NoStop}%
\bibitem [{\citenamefont {Palzer}\ \emph {et~al.}(2009)\citenamefont {Palzer},
  \citenamefont {Zipkes}, \citenamefont {Sias},\ and\ \citenamefont
  {K\"ohl}}]{Palzer2009}%
  \BibitemOpen
  \bibfield  {author} {\bibinfo {author} {\bibfnamefont {S.}~\bibnamefont
  {Palzer}}, \bibinfo {author} {\bibfnamefont {C.}~\bibnamefont {Zipkes}},
  \bibinfo {author} {\bibfnamefont {C.}~\bibnamefont {Sias}}, \ and\ \bibinfo
  {author} {\bibfnamefont {M.}~\bibnamefont {K\"ohl}},\ }\bibfield  {title}
  {\bibinfo {title} {\emph {Quantum Transport through a Tonks-Girardeau Gas}},\
  }\href {\doibase 10.1103/PhysRevLett.103.150601} {\bibfield  {journal}
  {\bibinfo  {journal} {Phys. Rev. Lett.}\ }\textbf {\bibinfo {volume} {103}},\
  \bibinfo {pages} {150601} (\bibinfo {year} {2009})}\BibitemShut {NoStop}%
\bibitem [{\citenamefont {Catani}\ \emph {et~al.}(2012)\citenamefont {Catani},
  \citenamefont {Lamporesi}, \citenamefont {Naik}, \citenamefont {Gring},
  \citenamefont {Inguscio}, \citenamefont {Minardi}, \citenamefont {Kantian},\
  and\ \citenamefont {Giamarchi}}]{Catani2012}%
  \BibitemOpen
  \bibfield  {author} {\bibinfo {author} {\bibfnamefont {J.}~\bibnamefont
  {Catani}}, \bibinfo {author} {\bibfnamefont {G.}~\bibnamefont {Lamporesi}},
  \bibinfo {author} {\bibfnamefont {D.}~\bibnamefont {Naik}}, \bibinfo {author}
  {\bibfnamefont {M.}~\bibnamefont {Gring}}, \bibinfo {author} {\bibfnamefont
  {M.}~\bibnamefont {Inguscio}}, \bibinfo {author} {\bibfnamefont
  {F.}~\bibnamefont {Minardi}}, \bibinfo {author} {\bibfnamefont
  {A.}~\bibnamefont {Kantian}}, \ and\ \bibinfo {author} {\bibfnamefont
  {T.}~\bibnamefont {Giamarchi}},\ }\bibfield  {title} {\bibinfo {title} {\emph
  {Quantum dynamics of impurities in a one-dimensional Bose gas}},\ }\href
  {\doibase 10.1103/PhysRevA.85.023623} {\bibfield  {journal} {\bibinfo
  {journal} {Phys. Rev. A}\ }\textbf {\bibinfo {volume} {85}},\ \bibinfo
  {pages} {023623} (\bibinfo {year} {2012})}\BibitemShut {NoStop}%
\bibitem [{\citenamefont {Spethmann}\ \emph {et~al.}(2012)\citenamefont
  {Spethmann}, \citenamefont {Kindermann}, \citenamefont {John}, \citenamefont
  {Weber}, \citenamefont {Meschede},\ and\ \citenamefont
  {Widera}}]{Spethmann2012}%
  \BibitemOpen
  \bibfield  {author} {\bibinfo {author} {\bibfnamefont {N.}~\bibnamefont
  {Spethmann}}, \bibinfo {author} {\bibfnamefont {F.}~\bibnamefont
  {Kindermann}}, \bibinfo {author} {\bibfnamefont {S.}~\bibnamefont {John}},
  \bibinfo {author} {\bibfnamefont {C.}~\bibnamefont {Weber}}, \bibinfo
  {author} {\bibfnamefont {D.}~\bibnamefont {Meschede}}, \ and\ \bibinfo
  {author} {\bibfnamefont {A.}~\bibnamefont {Widera}},\ }\bibfield  {title}
  {\bibinfo {title} {\emph {Dynamics of Single Neutral Impurity Atoms Immersed
  in an Ultracold Gas}},\ }\href {\doibase 10.1103/PhysRevLett.109.235301}
  {\bibfield  {journal} {\bibinfo  {journal} {Phys. Rev. Lett.}\ }\textbf
  {\bibinfo {volume} {109}},\ \bibinfo {pages} {235301} (\bibinfo {year}
  {2012})}\BibitemShut {NoStop}%
\bibitem [{\citenamefont {Rath}\ and\ \citenamefont
  {Schmidt}(2013)}]{Rath2013}%
  \BibitemOpen
  \bibfield  {author} {\bibinfo {author} {\bibfnamefont {S.~P.}\ \bibnamefont
  {Rath}}\ and\ \bibinfo {author} {\bibfnamefont {R.}~\bibnamefont {Schmidt}},\
  }\bibfield  {title} {\bibinfo {title} {\emph {Field-theoretical study of the
  Bose polaron}},\ }\href {\doibase 10.1103/PhysRevA.88.053632} {\bibfield
  {journal} {\bibinfo  {journal} {Phys. Rev. A}\ }\textbf {\bibinfo {volume}
  {88}},\ \bibinfo {pages} {053632} (\bibinfo {year} {2013})}\BibitemShut
  {NoStop}%
\bibitem [{\citenamefont {{Fukuhara}}\ \emph {et~al.}(2013)\citenamefont
  {{Fukuhara}}, \citenamefont {{Kantian}}, \citenamefont {{Endres}},
  \citenamefont {{Cheneau}}, \citenamefont {{Schau{\ss}}}, \citenamefont
  {{Hild}}, \citenamefont {{Bellem}}, \citenamefont {{Schollw{\"o}ck}},
  \citenamefont {{Giamarchi}}, \citenamefont {{Gross}}, \citenamefont
  {{Bloch}},\ and\ \citenamefont {{Kuhr}}}]{Fukuhara2013}%
  \BibitemOpen
  \bibfield  {author} {\bibinfo {author} {\bibfnamefont {T.}~\bibnamefont
  {{Fukuhara}}}, \bibinfo {author} {\bibfnamefont {A.}~\bibnamefont
  {{Kantian}}}, \bibinfo {author} {\bibfnamefont {M.}~\bibnamefont {{Endres}}},
  \bibinfo {author} {\bibfnamefont {M.}~\bibnamefont {{Cheneau}}}, \bibinfo
  {author} {\bibfnamefont {P.}~\bibnamefont {{Schau{\ss}}}}, \bibinfo {author}
  {\bibfnamefont {S.}~\bibnamefont {{Hild}}}, \bibinfo {author} {\bibfnamefont
  {D.}~\bibnamefont {{Bellem}}}, \bibinfo {author} {\bibfnamefont
  {U.}~\bibnamefont {{Schollw{\"o}ck}}}, \bibinfo {author} {\bibfnamefont
  {T.}~\bibnamefont {{Giamarchi}}}, \bibinfo {author} {\bibfnamefont
  {C.}~\bibnamefont {{Gross}}}, \bibinfo {author} {\bibfnamefont
  {I.}~\bibnamefont {{Bloch}}}, \ and\ \bibinfo {author} {\bibfnamefont
  {S.}~\bibnamefont {{Kuhr}}},\ }\bibfield  {title} {\bibinfo {title} {\emph
  {{Quantum dynamics of a mobile spin impurity}}},\ }\href {\doibase
  10.1038/nphys2561} {\bibfield  {journal} {\bibinfo  {journal} {Nature
  Physics}\ }\textbf {\bibinfo {volume} {9}},\ \bibinfo {pages} {235} (\bibinfo
  {year} {2013})}\BibitemShut {NoStop}%
\bibitem [{\citenamefont {Shashi}\ \emph {et~al.}(2014)\citenamefont {Shashi},
  \citenamefont {Grusdt}, \citenamefont {Abanin},\ and\ \citenamefont
  {Demler}}]{Shashi2014}%
  \BibitemOpen
  \bibfield  {author} {\bibinfo {author} {\bibfnamefont {A.}~\bibnamefont
  {Shashi}}, \bibinfo {author} {\bibfnamefont {F.}~\bibnamefont {Grusdt}},
  \bibinfo {author} {\bibfnamefont {D.~A.}\ \bibnamefont {Abanin}}, \ and\
  \bibinfo {author} {\bibfnamefont {E.}~\bibnamefont {Demler}},\ }\bibfield
  {title} {\bibinfo {title} {\emph {Radio-frequency spectroscopy of polarons in
  ultracold Bose gases}},\ }\href {\doibase 10.1103/PhysRevA.89.053617}
  {\bibfield  {journal} {\bibinfo  {journal} {Phys. Rev. A}\ }\textbf {\bibinfo
  {volume} {89}},\ \bibinfo {pages} {053617} (\bibinfo {year}
  {2014})}\BibitemShut {NoStop}%
\bibitem [{\citenamefont {Benjamin}\ and\ \citenamefont
  {Demler}(2014)}]{Benjamin2014}%
  \BibitemOpen
  \bibfield  {author} {\bibinfo {author} {\bibfnamefont {D.}~\bibnamefont
  {Benjamin}}\ and\ \bibinfo {author} {\bibfnamefont {E.}~\bibnamefont
  {Demler}},\ }\bibfield  {title} {\bibinfo {title} {\emph {Variational polaron
  method for Bose-Bose mixtures}},\ }\href {\doibase
  10.1103/PhysRevA.89.033615} {\bibfield  {journal} {\bibinfo  {journal} {Phys.
  Rev. A}\ }\textbf {\bibinfo {volume} {89}},\ \bibinfo {pages} {033615}
  (\bibinfo {year} {2014})}\BibitemShut {NoStop}%
\bibitem [{\citenamefont {{Grusdt}}\ \emph
  {et~al.}(2014{\natexlab{a}})\citenamefont {{Grusdt}}, \citenamefont
  {{Shashi}}, \citenamefont {{Abanin}},\ and\ \citenamefont
  {{Demler}}}]{Grusdt2014a}%
  \BibitemOpen
  \bibfield  {author} {\bibinfo {author} {\bibfnamefont {F.}~\bibnamefont
  {{Grusdt}}}, \bibinfo {author} {\bibfnamefont {A.}~\bibnamefont {{Shashi}}},
  \bibinfo {author} {\bibfnamefont {D.}~\bibnamefont {{Abanin}}}, \ and\
  \bibinfo {author} {\bibfnamefont {E.}~\bibnamefont {{Demler}}},\ }\bibfield
  {title} {\bibinfo {title} {\emph {{Bloch oscillations of bosonic lattice
  polarons}}},\ }\href@noop {} {\  (\bibinfo {year} {2014}{\natexlab{a}})},\
  \Eprint {http://arxiv.org/abs/1410.1513} {arXiv:1410.1513} \BibitemShut
  {NoStop}%
\bibitem [{\citenamefont {{Grusdt}}\ \emph
  {et~al.}(2014{\natexlab{b}})\citenamefont {{Grusdt}}, \citenamefont
  {{Shchadilova}}, \citenamefont {{Rubtsov}},\ and\ \citenamefont
  {{Demler}}}]{Grusdt2014b}%
  \BibitemOpen
  \bibfield  {author} {\bibinfo {author} {\bibfnamefont {F.}~\bibnamefont
  {{Grusdt}}}, \bibinfo {author} {\bibfnamefont {Y.~E.}\ \bibnamefont
  {{Shchadilova}}}, \bibinfo {author} {\bibfnamefont {A.~N.}\ \bibnamefont
  {{Rubtsov}}}, \ and\ \bibinfo {author} {\bibfnamefont {E.}~\bibnamefont
  {{Demler}}},\ }\bibfield  {title} {\bibinfo {title} {\emph {{Renormalization
  group approach to the Fr\"ohlich polaron model: application to impurity-BEC
  problem}}},\ }\href@noop {} {\  (\bibinfo {year} {2014}{\natexlab{b}})},\
  \Eprint {http://arxiv.org/abs/1410.2203} {arXiv:1410.2203} \BibitemShut
  {NoStop}%
\bibitem [{\citenamefont {Christensen}\ \emph
  {et~al.}(2015{\natexlab{a}})\citenamefont {Christensen}, \citenamefont
  {Levinsen},\ and\ \citenamefont {Bruun}}]{Chris2015}%
  \BibitemOpen
  \bibfield  {author} {\bibinfo {author} {\bibfnamefont {R.~S.}\ \bibnamefont
  {Christensen}}, \bibinfo {author} {\bibfnamefont {J.}~\bibnamefont
  {Levinsen}}, \ and\ \bibinfo {author} {\bibfnamefont {G.~M.}\ \bibnamefont
  {Bruun}},\ }\bibfield  {title} {\bibinfo {title} {\emph {Quasiparticle
  Properties of a Mobile Impurity in a Bose-Einstein Condensate}},\ }\href
  {\doibase 10.1103/PhysRevLett.115.160401} {\bibfield  {journal} {\bibinfo
  {journal} {Phys. Rev. Lett.}\ }\textbf {\bibinfo {volume} {115}},\ \bibinfo
  {pages} {160401} (\bibinfo {year} {2015}{\natexlab{a}})}\BibitemShut
  {NoStop}%
\bibitem [{\citenamefont {Levinsen}\ \emph {et~al.}(2015)\citenamefont
  {Levinsen}, \citenamefont {Parish},\ and\ \citenamefont
  {Bruun}}]{Levinsen2015}%
  \BibitemOpen
  \bibfield  {author} {\bibinfo {author} {\bibfnamefont {J.}~\bibnamefont
  {Levinsen}}, \bibinfo {author} {\bibfnamefont {M.~M.}\ \bibnamefont
  {Parish}}, \ and\ \bibinfo {author} {\bibfnamefont {G.~M.}\ \bibnamefont
  {Bruun}},\ }\bibfield  {title} {\bibinfo {title} {\emph {Impurity in a
  Bose-Einstein Condensate and the Efimov Effect}},\ }\href {\doibase
  10.1103/PhysRevLett.115.125302} {\bibfield  {journal} {\bibinfo  {journal}
  {Phys. Rev. Lett.}\ }\textbf {\bibinfo {volume} {115}},\ \bibinfo {pages}
  {125302} (\bibinfo {year} {2015})}\BibitemShut {NoStop}%
\bibitem [{\citenamefont {Ardila}\ and\ \citenamefont
  {Giorgini}(2015)}]{Ardila2015}%
  \BibitemOpen
  \bibfield  {author} {\bibinfo {author} {\bibfnamefont {L.~A.~P.}\
  \bibnamefont {Ardila}}\ and\ \bibinfo {author} {\bibfnamefont
  {S.}~\bibnamefont {Giorgini}},\ }\bibfield  {title} {\bibinfo {title} {\emph
  {Impurity in a Bose-Einstein condensate: Study of the attractive and
  repulsive branch using quantum Monte Carlo methods}},\ }\href {\doibase
  10.1103/PhysRevA.92.033612} {\bibfield  {journal} {\bibinfo  {journal} {Phys.
  Rev. A}\ }\textbf {\bibinfo {volume} {92}},\ \bibinfo {pages} {033612}
  (\bibinfo {year} {2015})}\BibitemShut {NoStop}%
\bibitem [{\citenamefont {Volosniev}\ \emph {et~al.}(2015)\citenamefont
  {Volosniev}, \citenamefont {Hammer},\ and\ \citenamefont
  {Zinner}}]{Volosniev2015}%
  \BibitemOpen
  \bibfield  {author} {\bibinfo {author} {\bibfnamefont {A.~G.}\ \bibnamefont
  {Volosniev}}, \bibinfo {author} {\bibfnamefont {H.-W.}\ \bibnamefont
  {Hammer}}, \ and\ \bibinfo {author} {\bibfnamefont {N.~T.}\ \bibnamefont
  {Zinner}},\ }\bibfield  {title} {\bibinfo {title} {\emph {Real-time dynamics
  of an impurity in an ideal Bose gas in a trap}},\ }\href {\doibase
  10.1103/PhysRevA.92.023623} {\bibfield  {journal} {\bibinfo  {journal} {Phys.
  Rev. A}\ }\textbf {\bibinfo {volume} {92}},\ \bibinfo {pages} {023623}
  (\bibinfo {year} {2015})}\BibitemShut {NoStop}%
\bibitem [{\citenamefont {Grusdt}\ and\ \citenamefont
  {Demler}(2016)}]{Grusdt2016}%
  \BibitemOpen
  \bibfield  {author} {\bibinfo {author} {\bibfnamefont {F.}~\bibnamefont
  {Grusdt}}\ and\ \bibinfo {author} {\bibfnamefont {E.}~\bibnamefont
  {Demler}},\ }\bibfield  {title} {\bibinfo {title} {\emph {New theoretical
  approaches to Bose polarons}},\ }\href {https://arxiv.org/pdf/1510.04934.pdf}
  {\bibfield  {journal} {\bibinfo  {journal} {arxiv}\ }\textbf {\bibinfo
  {volume} {1510.04934}} (\bibinfo {year} {2016})}\BibitemShut {NoStop}%
\bibitem [{\citenamefont {Grusdt}\ and\ \citenamefont
  {Fleischhauer}(2016)}]{Grusdt2016Feb}%
  \BibitemOpen
  \bibfield  {author} {\bibinfo {author} {\bibfnamefont {F.}~\bibnamefont
  {Grusdt}}\ and\ \bibinfo {author} {\bibfnamefont {M.}~\bibnamefont
  {Fleischhauer}},\ }\bibfield  {title} {\bibinfo {title} {\emph {Tunable
  Polarons of Slow-Light Polaritons in a Two-Dimensional Bose-Einstein
  Condensate}},\ }\href {\doibase 10.1103/PhysRevLett.116.053602} {\bibfield
  {journal} {\bibinfo  {journal} {Phys. Rev. Lett.}\ }\textbf {\bibinfo
  {volume} {116}},\ \bibinfo {pages} {053602} (\bibinfo {year}
  {2016})}\BibitemShut {NoStop}%
\bibitem [{\citenamefont {Shchadilova}\ \emph
  {et~al.}(2016{\natexlab{a}})\citenamefont {Shchadilova}, \citenamefont
  {Schmidt}, \citenamefont {Grusdt},\ and\ \citenamefont
  {Demler}}]{Shchadilova2016}%
  \BibitemOpen
  \bibfield  {author} {\bibinfo {author} {\bibfnamefont {Y.~E.}\ \bibnamefont
  {Shchadilova}}, \bibinfo {author} {\bibfnamefont {R.}~\bibnamefont
  {Schmidt}}, \bibinfo {author} {\bibfnamefont {F.}~\bibnamefont {Grusdt}}, \
  and\ \bibinfo {author} {\bibfnamefont {E.}~\bibnamefont {Demler}},\
  }\bibfield  {title} {\bibinfo {title} {\emph {Quantum Dynamics of Ultracold
  Bose Polarons}},\ }\href {\doibase 10.1103/PhysRevLett.117.113002} {\bibfield
   {journal} {\bibinfo  {journal} {Phys. Rev. Lett.}\ }\textbf {\bibinfo
  {volume} {117}},\ \bibinfo {pages} {113002} (\bibinfo {year}
  {2016}{\natexlab{a}})}\BibitemShut {NoStop}%
\bibitem [{\citenamefont {Shchadilova}\ \emph
  {et~al.}(2016{\natexlab{b}})\citenamefont {Shchadilova}, \citenamefont
  {Grusdt}, \citenamefont {Rubtsov},\ and\ \citenamefont
  {Demler}}]{Shchadilova2016b}%
  \BibitemOpen
  \bibfield  {author} {\bibinfo {author} {\bibfnamefont {Y.~E.}\ \bibnamefont
  {Shchadilova}}, \bibinfo {author} {\bibfnamefont {F.}~\bibnamefont {Grusdt}},
  \bibinfo {author} {\bibfnamefont {A.~N.}\ \bibnamefont {Rubtsov}}, \ and\
  \bibinfo {author} {\bibfnamefont {E.}~\bibnamefont {Demler}},\ }\bibfield
  {title} {\bibinfo {title} {\emph {Polaronic mass renormalization of
  impurities in Bose-Einstein condensates: Correlated Gaussian-wave-function
  approach}},\ }\href {\doibase 10.1103/PhysRevA.93.043606} {\bibfield
  {journal} {\bibinfo  {journal} {Phys. Rev. A}\ }\textbf {\bibinfo {volume}
  {93}},\ \bibinfo {pages} {043606} (\bibinfo {year}
  {2016}{\natexlab{b}})}\BibitemShut {NoStop}%
\bibitem [{\citenamefont {Castelnovo}\ \emph {et~al.}(2016)\citenamefont
  {Castelnovo}, \citenamefont {Caux},\ and\ \citenamefont
  {Simon}}]{Castelnovo2016}%
  \BibitemOpen
  \bibfield  {author} {\bibinfo {author} {\bibfnamefont {C.}~\bibnamefont
  {Castelnovo}}, \bibinfo {author} {\bibfnamefont {J.-S.}\ \bibnamefont
  {Caux}}, \ and\ \bibinfo {author} {\bibfnamefont {S.~H.}\ \bibnamefont
  {Simon}},\ }\bibfield  {title} {\bibinfo {title} {\emph {Driven impurity in
  an ultracold one-dimensional Bose gas with intermediate interaction
  strength}},\ }\href {\doibase 10.1103/PhysRevA.93.013613} {\bibfield
  {journal} {\bibinfo  {journal} {Phys. Rev. A}\ }\textbf {\bibinfo {volume}
  {93}},\ \bibinfo {pages} {013613} (\bibinfo {year} {2016})}\BibitemShut
  {NoStop}%
\bibitem [{\citenamefont {Ardila}\ and\ \citenamefont
  {Giorgini}(2016)}]{Ardila2016}%
  \BibitemOpen
  \bibfield  {author} {\bibinfo {author} {\bibfnamefont {L.~A.~P.}\
  \bibnamefont {Ardila}}\ and\ \bibinfo {author} {\bibfnamefont
  {S.}~\bibnamefont {Giorgini}},\ }\bibfield  {title} {\bibinfo {title} {\emph
  {Bose polaron problem: Effect of mass imbalance on binding energy}},\ }\href
  {\doibase 10.1103/PhysRevA.94.063640} {\bibfield  {journal} {\bibinfo
  {journal} {Phys. Rev. A}\ }\textbf {\bibinfo {volume} {94}},\ \bibinfo
  {pages} {063640} (\bibinfo {year} {2016})}\BibitemShut {NoStop}%
\bibitem [{\citenamefont {Robinson}\ \emph {et~al.}(2016)\citenamefont
  {Robinson}, \citenamefont {Caux},\ and\ \citenamefont
  {Konik}}]{Robinson2016}%
  \BibitemOpen
  \bibfield  {author} {\bibinfo {author} {\bibfnamefont {N.~J.}\ \bibnamefont
  {Robinson}}, \bibinfo {author} {\bibfnamefont {J.-S.}\ \bibnamefont {Caux}},
  \ and\ \bibinfo {author} {\bibfnamefont {R.~M.}\ \bibnamefont {Konik}},\
  }\bibfield  {title} {\bibinfo {title} {\emph {Motion of a Distinguishable
  Impurity in the Bose Gas: Arrested Expansion Without a Lattice and Impurity
  Snaking}},\ }\href {\doibase 10.1103/PhysRevLett.116.145302} {\bibfield
  {journal} {\bibinfo  {journal} {Phys. Rev. Lett.}\ }\textbf {\bibinfo
  {volume} {116}},\ \bibinfo {pages} {145302} (\bibinfo {year}
  {2016})}\BibitemShut {NoStop}%
\bibitem [{\citenamefont {J\o{}rgensen}\ \emph {et~al.}(2016)\citenamefont
  {J\o{}rgensen}, \citenamefont {Wacker}, \citenamefont {Skalmstang},
  \citenamefont {Parish}, \citenamefont {Levinsen}, \citenamefont
  {Christensen}, \citenamefont {Bruun},\ and\ \citenamefont {Arlt}}]{Jor2016}%
  \BibitemOpen
  \bibfield  {author} {\bibinfo {author} {\bibfnamefont {N.~B.}\ \bibnamefont
  {J\o{}rgensen}}, \bibinfo {author} {\bibfnamefont {L.}~\bibnamefont
  {Wacker}}, \bibinfo {author} {\bibfnamefont {K.~T.}\ \bibnamefont
  {Skalmstang}}, \bibinfo {author} {\bibfnamefont {M.~M.}\ \bibnamefont
  {Parish}}, \bibinfo {author} {\bibfnamefont {J.}~\bibnamefont {Levinsen}},
  \bibinfo {author} {\bibfnamefont {R.~S.}\ \bibnamefont {Christensen}},
  \bibinfo {author} {\bibfnamefont {G.~M.}\ \bibnamefont {Bruun}}, \ and\
  \bibinfo {author} {\bibfnamefont {J.~J.}\ \bibnamefont {Arlt}},\ }\bibfield
  {title} {\bibinfo {title} {\emph {Observation of Attractive and Repulsive
  Polarons in a Bose-Einstein Condensate}},\ }\href {\doibase
  10.1103/PhysRevLett.117.055302} {\bibfield  {journal} {\bibinfo  {journal}
  {Phys. Rev. Lett.}\ }\textbf {\bibinfo {volume} {117}},\ \bibinfo {pages}
  {055302} (\bibinfo {year} {2016})}\BibitemShut {NoStop}%
\bibitem [{\citenamefont {Hu}\ \emph {et~al.}(2016)\citenamefont {Hu},
  \citenamefont {Van~de Graaff}, \citenamefont {Kedar}, \citenamefont {Corson},
  \citenamefont {Cornell},\ and\ \citenamefont {Jin}}]{Hu2016}%
  \BibitemOpen
  \bibfield  {author} {\bibinfo {author} {\bibfnamefont {M.-G.}\ \bibnamefont
  {Hu}}, \bibinfo {author} {\bibfnamefont {M.~J.}\ \bibnamefont {Van~de
  Graaff}}, \bibinfo {author} {\bibfnamefont {D.}~\bibnamefont {Kedar}},
  \bibinfo {author} {\bibfnamefont {J.~P.}\ \bibnamefont {Corson}}, \bibinfo
  {author} {\bibfnamefont {E.~A.}\ \bibnamefont {Cornell}}, \ and\ \bibinfo
  {author} {\bibfnamefont {D.~S.}\ \bibnamefont {Jin}},\ }\bibfield  {title}
  {\bibinfo {title} {\emph {Bose Polarons in the Strongly Interacting
  Regime}},\ }\href {\doibase 10.1103/PhysRevLett.117.055301} {\bibfield
  {journal} {\bibinfo  {journal} {Phys. Rev. Lett.}\ }\textbf {\bibinfo
  {volume} {117}},\ \bibinfo {pages} {055301} (\bibinfo {year}
  {2016})}\BibitemShut {NoStop}%
\bibitem [{\citenamefont {Rentrop}\ \emph {et~al.}(2016)\citenamefont
  {Rentrop}, \citenamefont {Trautmann}, \citenamefont {Olivares}, \citenamefont
  {Jendrzejewski}, \citenamefont {Komnik},\ and\ \citenamefont
  {Oberthaler}}]{Rentrop2016}%
  \BibitemOpen
  \bibfield  {author} {\bibinfo {author} {\bibfnamefont {T.}~\bibnamefont
  {Rentrop}}, \bibinfo {author} {\bibfnamefont {A.}~\bibnamefont {Trautmann}},
  \bibinfo {author} {\bibfnamefont {F.~A.}\ \bibnamefont {Olivares}}, \bibinfo
  {author} {\bibfnamefont {F.}~\bibnamefont {Jendrzejewski}}, \bibinfo {author}
  {\bibfnamefont {A.}~\bibnamefont {Komnik}}, \ and\ \bibinfo {author}
  {\bibfnamefont {M.~K.}\ \bibnamefont {Oberthaler}},\ }\bibfield  {title}
  {\bibinfo {title} {\emph {Observation of the Phononic Lamb Shift with a
  Synthetic Vacuum}},\ }\href {\doibase 10.1103/PhysRevX.6.041041} {\bibfield
  {journal} {\bibinfo  {journal} {Phys. Rev. X}\ }\textbf {\bibinfo {volume}
  {6}},\ \bibinfo {pages} {041041} (\bibinfo {year} {2016})}\BibitemShut
  {NoStop}%
\bibitem [{\citenamefont {Gardiner}\ and\ \citenamefont
  {Zoller}(2004)}]{GardinerBook}%
  \BibitemOpen
  \bibfield  {author} {\bibinfo {author} {\bibfnamefont {C.}~\bibnamefont
  {Gardiner}}\ and\ \bibinfo {author} {\bibfnamefont {P.}~\bibnamefont
  {Zoller}},\ }\href@noop {} {\emph {\bibinfo {title} {Quantum Noise: A
  Handbook of Markovian and Non-Markovian Quantum Stochastic Methods with
  Applications to Quantum Optics}}},\ Springer Series in Synergetics\ (\bibinfo
   {publisher} {Springer},\ \bibinfo {address} {Berlin},\ \bibinfo {year}
  {2004})\BibitemShut {NoStop}%
\bibitem [{\citenamefont {Breuer}\ and\ \citenamefont
  {Petruccione}(2007)}]{BreuerBook}%
  \BibitemOpen
  \bibfield  {author} {\bibinfo {author} {\bibfnamefont {H.}~\bibnamefont
  {Breuer}}\ and\ \bibinfo {author} {\bibfnamefont {F.}~\bibnamefont
  {Petruccione}},\ }\href {http://books.google.es/books?id=DkcJPwAACAAJ} {\emph
  {\bibinfo {title} {The Theory of Open Quantum Systems}}}\ (\bibinfo
  {publisher} {OUP},\ \bibinfo {address} {Oxford},\ \bibinfo {year}
  {2007})\BibitemShut {NoStop}%
\bibitem [{\citenamefont {Schlosshauer}(2007)}]{SchlosshauerBook}%
  \BibitemOpen
  \bibfield  {author} {\bibinfo {author} {\bibfnamefont {M.}~\bibnamefont
  {Schlosshauer}},\ }\href
  {http://www.springer.com/physics/quantum+physics/book/978-3-540-35773-5}
  {\emph {\bibinfo {title} {Decoherence and the Quantum-To-Classical
  Transition}}},\ The Frontiers Collection\ (\bibinfo  {publisher} {Springer},\
  \bibinfo {year} {2007})\BibitemShut {NoStop}%
\bibitem [{\citenamefont {Schlosshauer}(2005)}]{Schlosshauer2005}%
  \BibitemOpen
  \bibfield  {author} {\bibinfo {author} {\bibfnamefont {M.}~\bibnamefont
  {Schlosshauer}},\ }\bibfield  {title} {\bibinfo {title} {\emph {Decoherence,
  the measurement problem, and interpretations of quantum mechanics}},\ }\href
  {\doibase 10.1103/RevModPhys.76.1267} {\bibfield  {journal} {\bibinfo
  {journal} {Rev. Mod. Phys.}\ }\textbf {\bibinfo {volume} {76}},\ \bibinfo
  {pages} {1267} (\bibinfo {year} {2005})}\BibitemShut {NoStop}%
\bibitem [{\citenamefont {Zurek}(2003)}]{Zurek2003}%
  \BibitemOpen
  \bibfield  {author} {\bibinfo {author} {\bibfnamefont {W.~H.}\ \bibnamefont
  {Zurek}},\ }\bibfield  {title} {\bibinfo {title} {\emph {Decoherence,
  einselection, and the quantum origins of the classical}},\ }\href {\doibase
  10.1103/RevModPhys.75.715} {\bibfield  {journal} {\bibinfo  {journal} {Rev.
  Mod. Phys.}\ }\textbf {\bibinfo {volume} {75}},\ \bibinfo {pages} {715}
  (\bibinfo {year} {2003})}\BibitemShut {NoStop}%
\bibitem [{\citenamefont {Caldeira}\ and\ \citenamefont
  {Leggett}(1983{\natexlab{a}})}]{Caldeira1983a}%
  \BibitemOpen
  \bibfield  {author} {\bibinfo {author} {\bibfnamefont {A.}~\bibnamefont
  {Caldeira}}\ and\ \bibinfo {author} {\bibfnamefont {A.}~\bibnamefont
  {Leggett}},\ }\bibfield  {title} {\bibinfo {title} {\emph {Path integral
  approach to quantum Brownian motion}},\ }\href {\doibase
  10.1016/0378-4371(83)90013-4} {\bibfield  {journal} {\bibinfo  {journal}
  {Physica A: Statistical Mechanics and its Applications}\ }\textbf {\bibinfo
  {volume} {121}},\ \bibinfo {pages} {587 } (\bibinfo {year}
  {1983}{\natexlab{a}})}\BibitemShut {NoStop}%
\bibitem [{\citenamefont {Caldeira}\ and\ \citenamefont
  {Leggett}(1983{\natexlab{b}})}]{Caldeira1983b}%
  \BibitemOpen
  \bibfield  {author} {\bibinfo {author} {\bibfnamefont {A.}~\bibnamefont
  {Caldeira}}\ and\ \bibinfo {author} {\bibfnamefont {A.}~\bibnamefont
  {Leggett}},\ }\bibfield  {title} {\bibinfo {title} {\emph {Quantum tunnelling
  in a dissipative system}},\ }\href {\doibase 10.1016/0003-4916(83)90202-6}
  {\bibfield  {journal} {\bibinfo  {journal} {Annals of Physics}\ }\textbf
  {\bibinfo {volume} {149}},\ \bibinfo {pages} {374 } (\bibinfo {year}
  {1983}{\natexlab{b}})}\BibitemShut {NoStop}%
\bibitem [{\citenamefont {Efimkin}\ \emph {et~al.}(2016)\citenamefont
  {Efimkin}, \citenamefont {Hofmann},\ and\ \citenamefont
  {Galitski}}]{Efimikin2013}%
  \BibitemOpen
  \bibfield  {author} {\bibinfo {author} {\bibfnamefont {D.~K.}\ \bibnamefont
  {Efimkin}}, \bibinfo {author} {\bibfnamefont {J.}~\bibnamefont {Hofmann}}, \
  and\ \bibinfo {author} {\bibfnamefont {V.}~\bibnamefont {Galitski}},\
  }\bibfield  {title} {\bibinfo {title} {\emph {Non-Markovian Quantum Friction
  of Bright Solitons in Superfluids}},\ }\href {\doibase
  10.1103/PhysRevLett.116.225301} {\bibfield  {journal} {\bibinfo  {journal}
  {Phys. Rev. Lett.}\ }\textbf {\bibinfo {volume} {116}},\ \bibinfo {pages}
  {225301} (\bibinfo {year} {2016})}\BibitemShut {NoStop}%
\bibitem [{\citenamefont {Hilary M.~Hurst}(2016)}]{Hurst2016}%
  \BibitemOpen
  \bibfield  {author} {\bibinfo {author} {\bibfnamefont {I.~B. S. V.~G.}\
  \bibnamefont {Hilary M.~Hurst}, \bibfnamefont {Dmitry K.~Efimkin}},\
  }\bibfield  {title} {\bibinfo {title} {\emph {Kinetic theory of dark solitons
  with tunable friction}},\ }\href {https://arxiv.org/abs/1703.00809}
  {\bibfield  {journal} {\bibinfo  {journal} {arxiv}\ } (\bibinfo {year}
  {2016})}\BibitemShut {NoStop}%
\bibitem [{\citenamefont {Keser}\ and\ \citenamefont
  {Galitski}(2016)}]{Keser2016}%
  \BibitemOpen
  \bibfield  {author} {\bibinfo {author} {\bibfnamefont {A.~C.}\ \bibnamefont
  {Keser}}\ and\ \bibinfo {author} {\bibfnamefont {V.}~\bibnamefont
  {Galitski}},\ }\bibfield  {title} {\bibinfo {title} {\emph {Analogue
  Stochastic Gravity in Strongly-Interacting Bose-Einstein Condensates}},\
  }\href {https://arxiv.org/pdf/1612.08980.pdf} {\bibfield  {journal} {\bibinfo
   {journal} {arXiv:1612.08980}\ } (\bibinfo {year} {2016})}\BibitemShut
  {NoStop}%
\bibitem [{\citenamefont {Bonart}\ and\ \citenamefont
  {Cugliandolo}(2012)}]{Cugliandolo2012}%
  \BibitemOpen
  \bibfield  {author} {\bibinfo {author} {\bibfnamefont {J.}~\bibnamefont
  {Bonart}}\ and\ \bibinfo {author} {\bibfnamefont {L.~F.}\ \bibnamefont
  {Cugliandolo}},\ }\bibfield  {title} {\bibinfo {title} {\emph {From
  nonequilibrium quantum Brownian motion to impurity dynamics in
  one-dimensional quantum liquids}},\ }\href {\doibase
  10.1103/PhysRevA.86.023636} {\bibfield  {journal} {\bibinfo  {journal} {Phys.
  Rev. A}\ }\textbf {\bibinfo {volume} {86}},\ \bibinfo {pages} {023636}
  (\bibinfo {year} {2012})}\BibitemShut {NoStop}%
\bibitem [{\citenamefont {Bonart}\ and\ \citenamefont
  {Cugliandolo}(2013)}]{Bonart2013}%
  \BibitemOpen
  \bibfield  {author} {\bibinfo {author} {\bibfnamefont {J.}~\bibnamefont
  {Bonart}}\ and\ \bibinfo {author} {\bibfnamefont {L.~F.}\ \bibnamefont
  {Cugliandolo}},\ }\bibfield  {title} {\bibinfo {title} {\emph {Effective
  potential and polaronic mass shift in a trapped dynamical impurity Luttinger
  liquid system}},\ }\href {http://stacks.iop.org/0295-5075/101/i=1/a=16003}
  {\bibfield  {journal} {\bibinfo  {journal} {EPL (Europhysics Letters)}\
  }\textbf {\bibinfo {volume} {101}},\ \bibinfo {pages} {16003} (\bibinfo
  {year} {2013})}\BibitemShut {NoStop}%
\bibitem [{\citenamefont {Massignan}\ \emph {et~al.}(2015)\citenamefont
  {Massignan}, \citenamefont {Lampo}, \citenamefont {Wehr},\ and\ \citenamefont
  {Lewenstein}}]{Massignan2015}%
  \BibitemOpen
  \bibfield  {author} {\bibinfo {author} {\bibfnamefont {P.}~\bibnamefont
  {Massignan}}, \bibinfo {author} {\bibfnamefont {A.}~\bibnamefont {Lampo}},
  \bibinfo {author} {\bibfnamefont {J.}~\bibnamefont {Wehr}}, \ and\ \bibinfo
  {author} {\bibfnamefont {M.}~\bibnamefont {Lewenstein}},\ }\bibfield  {title}
  {\bibinfo {title} {\emph {Quantum Brownian motion with inhomogeneous damping
  and diffusion}},\ }\href {\doibase 10.1103/PhysRevA.91.033627} {\bibfield
  {journal} {\bibinfo  {journal} {Phys. Rev. A}\ }\textbf {\bibinfo {volume}
  {91}},\ \bibinfo {pages} {033627} (\bibinfo {year} {2015})}\BibitemShut
  {NoStop}%
\bibitem [{\citenamefont {Lampo}\ \emph {et~al.}(2016)\citenamefont {Lampo},
  \citenamefont {Lim}, \citenamefont {Wehr}, \citenamefont {Massignan},\ and\
  \citenamefont {Lewenstein}}]{Lampo2016}%
  \BibitemOpen
  \bibfield  {author} {\bibinfo {author} {\bibfnamefont {A.}~\bibnamefont
  {Lampo}}, \bibinfo {author} {\bibfnamefont {S.~H.}\ \bibnamefont {Lim}},
  \bibinfo {author} {\bibfnamefont {J.}~\bibnamefont {Wehr}}, \bibinfo {author}
  {\bibfnamefont {P.}~\bibnamefont {Massignan}}, \ and\ \bibinfo {author}
  {\bibfnamefont {M.}~\bibnamefont {Lewenstein}},\ }\bibfield  {title}
  {\bibinfo {title} {\emph {Lindblad model of quantum Brownian motion}},\
  }\href {\doibase 10.1103/PhysRevA.94.042123} {\bibfield  {journal} {\bibinfo
  {journal} {Phys. Rev. A}\ }\textbf {\bibinfo {volume} {94}},\ \bibinfo
  {pages} {042123} (\bibinfo {year} {2016})}\BibitemShut {NoStop}%
\bibitem [{\citenamefont {Pitaevskii}\ and\ \citenamefont
  {Stringari}(2003)}]{Pita-String}%
  \BibitemOpen
  \bibfield  {author} {\bibinfo {author} {\bibfnamefont {L.}~\bibnamefont
  {Pitaevskii}}\ and\ \bibinfo {author} {\bibfnamefont {S.}~\bibnamefont
  {Stringari}},\ }\href@noop {} {\emph {\bibinfo {title} {Bose-Einstein
  Condensation}}}\ (\bibinfo  {publisher} {Oxford University Press},\ \bibinfo
  {address} {Oxford},\ \bibinfo {year} {2003})\BibitemShut {NoStop}%
\bibitem [{\citenamefont {Canizares}\ and\ \citenamefont
  {Sols}(1994)}]{Sanchez1994}%
  \BibitemOpen
  \bibfield  {author} {\bibinfo {author} {\bibfnamefont {J.~S.}\ \bibnamefont
  {Canizares}}\ and\ \bibinfo {author} {\bibfnamefont {F.}~\bibnamefont
  {Sols}},\ }\bibfield  {title} {\bibinfo {title} {\emph {Translational
  symmetry and microscopic preparation in oscillator models of quantum
  dissipation}},\ }\href {\doibase 10.1016/0378-4371(94)90146-5} {\bibfield
  {journal} {\bibinfo  {journal} {Physica A: Statistical Mechanics and its
  Applications}\ }\textbf {\bibinfo {volume} {212}},\ \bibinfo {pages} {181 }
  (\bibinfo {year} {1994})}\BibitemShut {NoStop}%
\bibitem [{\citenamefont {Bruderer}\ \emph {et~al.}(2007)\citenamefont
  {Bruderer}, \citenamefont {Klein}, \citenamefont {Clark},\ and\ \citenamefont
  {Jaksch}}]{Bruderer2007}%
  \BibitemOpen
  \bibfield  {author} {\bibinfo {author} {\bibfnamefont {M.}~\bibnamefont
  {Bruderer}}, \bibinfo {author} {\bibfnamefont {A.}~\bibnamefont {Klein}},
  \bibinfo {author} {\bibfnamefont {S.~R.}\ \bibnamefont {Clark}}, \ and\
  \bibinfo {author} {\bibfnamefont {D.}~\bibnamefont {Jaksch}},\ }\bibfield
  {title} {\bibinfo {title} {\emph {Polaron physics in optical lattices}},\
  }\href {\doibase 10.1103/PhysRevA.76.011605} {\bibfield  {journal} {\bibinfo
  {journal} {Phys. Rev. A}\ }\textbf {\bibinfo {volume} {76}},\ \bibinfo
  {pages} {011605} (\bibinfo {year} {2007})}\BibitemShut {NoStop}%
\bibitem [{\citenamefont {Christensen}\ \emph
  {et~al.}(2015{\natexlab{b}})\citenamefont {Christensen}, \citenamefont
  {Levinsen},\ and\ \citenamefont {Bruun}}]{Brunn2015}%
  \BibitemOpen
  \bibfield  {author} {\bibinfo {author} {\bibfnamefont {R.~S.}\ \bibnamefont
  {Christensen}}, \bibinfo {author} {\bibfnamefont {J.}~\bibnamefont
  {Levinsen}}, \ and\ \bibinfo {author} {\bibfnamefont {G.~M.}\ \bibnamefont
  {Bruun}},\ }\bibfield  {title} {\bibinfo {title} {\emph {Quasiparticle
  Properties of a Mobile Impurity in a Bose-Einstein Condensate}},\ }\href
  {\doibase 10.1103/PhysRevLett.115.160401} {\bibfield  {journal} {\bibinfo
  {journal} {Phys. Rev. Lett.}\ }\textbf {\bibinfo {volume} {115}},\ \bibinfo
  {pages} {160401} (\bibinfo {year} {2015}{\natexlab{b}})}\BibitemShut
  {NoStop}%
\bibitem [{\citenamefont {Tempere}\ \emph {et~al.}(2009)\citenamefont
  {Tempere}, \citenamefont {Casteels}, \citenamefont {Oberthaler},
  \citenamefont {Knoop}, \citenamefont {Timmermans},\ and\ \citenamefont
  {Devreese}}]{Tempere2009}%
  \BibitemOpen
  \bibfield  {author} {\bibinfo {author} {\bibfnamefont {J.}~\bibnamefont
  {Tempere}}, \bibinfo {author} {\bibfnamefont {W.}~\bibnamefont {Casteels}},
  \bibinfo {author} {\bibfnamefont {M.}~\bibnamefont {Oberthaler}}, \bibinfo
  {author} {\bibfnamefont {S.}~\bibnamefont {Knoop}}, \bibinfo {author}
  {\bibfnamefont {E.}~\bibnamefont {Timmermans}}, \ and\ \bibinfo {author}
  {\bibfnamefont {J.}~\bibnamefont {Devreese}},\ }\bibfield  {title} {\bibinfo
  {title} {\emph {Feynman path-integral treatment of the BEC-impurity
  polaron}},\ }\href {\doibase 10.1103/PhysRevB.80.184504} {\bibfield
  {journal} {\bibinfo  {journal} {Phys. Rev. B}\ }\textbf {\bibinfo {volume}
  {80}},\ \bibinfo {pages} {184504} (\bibinfo {year} {2009})}\BibitemShut
  {NoStop}%
\bibitem [{\citenamefont {Casteels}\ and\ \citenamefont
  {Wouters}(2014)}]{Casteels2014}%
  \BibitemOpen
  \bibfield  {author} {\bibinfo {author} {\bibfnamefont {W.}~\bibnamefont
  {Casteels}}\ and\ \bibinfo {author} {\bibfnamefont {M.}~\bibnamefont
  {Wouters}},\ }\bibfield  {title} {\bibinfo {title} {\emph {Polaron formation
  in the vicinity of a narrow Feshbach resonance}},\ }\href {\doibase
  10.1103/PhysRevA.90.043602} {\bibfield  {journal} {\bibinfo  {journal} {Phys.
  Rev. A}\ }\textbf {\bibinfo {volume} {90}},\ \bibinfo {pages} {043602}
  (\bibinfo {year} {2014})}\BibitemShut {NoStop}%
\bibitem [{\citenamefont {Grusdt}\ \emph {et~al.}(2017)\citenamefont {Grusdt},
  \citenamefont {Astrakharchik},\ and\ \citenamefont {Demler}}]{Grusdt2017}%
  \BibitemOpen
  \bibfield  {author} {\bibinfo {author} {\bibfnamefont {F.}~\bibnamefont
  {Grusdt}}, \bibinfo {author} {\bibfnamefont {G.~E.}\ \bibnamefont
  {Astrakharchik}}, \ and\ \bibinfo {author} {\bibfnamefont {E.~A.}\
  \bibnamefont {Demler}},\ }\bibfield  {title} {\bibinfo {title} {\emph {Bose
  polarons in ultracold atoms in one dimension: beyond the Fr{\"o}hlich
  paradigm}},\ }\href {https://arxiv.org/pdf/1704.02606.pdf} {\bibfield
  {journal} {\bibinfo  {journal} {arXiv:1704.02606}\ } (\bibinfo {year}
  {2017})}\BibitemShut {NoStop}%
\bibitem [{\citenamefont {Peotta}\ \emph {et~al.}(2013)\citenamefont {Peotta},
  \citenamefont {Rossini}, \citenamefont {Polini}, \citenamefont {Minardi},\
  and\ \citenamefont {Fazio}}]{Peotta2013}%
  \BibitemOpen
  \bibfield  {author} {\bibinfo {author} {\bibfnamefont {S.}~\bibnamefont
  {Peotta}}, \bibinfo {author} {\bibfnamefont {D.}~\bibnamefont {Rossini}},
  \bibinfo {author} {\bibfnamefont {M.}~\bibnamefont {Polini}}, \bibinfo
  {author} {\bibfnamefont {F.}~\bibnamefont {Minardi}}, \ and\ \bibinfo
  {author} {\bibfnamefont {R.}~\bibnamefont {Fazio}},\ }\bibfield  {title}
  {\bibinfo {title} {\emph {Quantum Breathing of an Impurity in a
  One-Dimensional Bath of Interacting Bosons}},\ }\href {\doibase
  10.1103/PhysRevLett.110.015302} {\bibfield  {journal} {\bibinfo  {journal}
  {Phys. Rev. Lett.}\ }\textbf {\bibinfo {volume} {110}},\ \bibinfo {pages}
  {015302} (\bibinfo {year} {2013})}\BibitemShut {NoStop}%
\bibitem [{\citenamefont {Ford}\ and\ \citenamefont
  {O'Connell}(1991)}]{Ford1991}%
  \BibitemOpen
  \bibfield  {author} {\bibinfo {author} {\bibfnamefont {G.~W.}\ \bibnamefont
  {Ford}}\ and\ \bibinfo {author} {\bibfnamefont {R.~F.}\ \bibnamefont
  {O'Connell}},\ }\bibfield  {title} {\bibinfo {title} {\emph {Radiation
  reaction in electrodynamics and the elimination of runaway solutions}},\
  }\href {\doibase 10.1016/0375-9601(91)90054-C} {\bibfield  {journal}
  {\bibinfo  {journal} {Physics Letters A}\ }\textbf {\bibinfo {volume}
  {157}},\ \bibinfo {pages} {217} (\bibinfo {year} {1991})}\BibitemShut
  {NoStop}%
\bibitem [{\citenamefont {Wang}\ and\ \citenamefont {Zhan}(2015)}]{Wang2015}%
  \BibitemOpen
  \bibfield  {author} {\bibinfo {author} {\bibfnamefont {Q.}~\bibnamefont
  {Wang}}\ and\ \bibinfo {author} {\bibfnamefont {H.}~\bibnamefont {Zhan}},\
  }\bibfield  {title} {\bibinfo {title} {\emph {On different numerical inverse
  Laplace methods for solute transport problems}},\ }\href {\doibase
  10.1016/0375-9601(91)90054-C} {\bibfield  {journal} {\bibinfo  {journal}
  {Advances in Water Resources}\ }\textbf {\bibinfo {volume} {75}},\ \bibinfo
  {pages} {80} (\bibinfo {year} {2015})}\BibitemShut {NoStop}%
\bibitem [{\citenamefont {Guarnieri}\ \emph {et~al.}(2016)\citenamefont
  {Guarnieri}, \citenamefont {Uchiyama},\ and\ \citenamefont
  {Vacchini}}]{Guarnieri2016}%
  \BibitemOpen
  \bibfield  {author} {\bibinfo {author} {\bibfnamefont {G.}~\bibnamefont
  {Guarnieri}}, \bibinfo {author} {\bibfnamefont {C.}~\bibnamefont {Uchiyama}},
  \ and\ \bibinfo {author} {\bibfnamefont {B.}~\bibnamefont {Vacchini}},\
  }\bibfield  {title} {\bibinfo {title} {\emph {Energy backflow and
  non-Markovian dynamics}},\ }\href {\doibase 10.1103/PhysRevA.93.012118}
  {\bibfield  {journal} {\bibinfo  {journal} {Phys. Rev. A}\ }\textbf {\bibinfo
  {volume} {93}},\ \bibinfo {pages} {012118} (\bibinfo {year}
  {2016})}\BibitemShut {NoStop}%
\bibitem [{\citenamefont {Lim}\ \emph {et~al.}(2017)\citenamefont {Lim},
  \citenamefont {Wehr}, \citenamefont {Lampo}, \citenamefont
  {Garc\'{i}a-March},\ and\ \citenamefont {Lewenstein}}]{Lim2017}%
  \BibitemOpen
  \bibfield  {author} {\bibinfo {author} {\bibfnamefont {S.~H.}\ \bibnamefont
  {Lim}}, \bibinfo {author} {\bibfnamefont {J.}~\bibnamefont {Wehr}}, \bibinfo
  {author} {\bibfnamefont {A.}~\bibnamefont {Lampo}}, \bibinfo {author}
  {\bibfnamefont {M.~A.}\ \bibnamefont {Garc\'{i}a-March}}, \ and\ \bibinfo
  {author} {\bibfnamefont {M.}~\bibnamefont {Lewenstein}},\ }\bibfield  {title}
  {\bibinfo {title} {\emph {On the Small Mass Limit of Quantum Brownian Motion
  with Inhomogeneous Damping and Diffusion}},\ }\href
  {https://arxiv.org/pdf/1708.03685.pdf} {\bibfield  {journal} {\bibinfo
  {journal} {arXiv:1704.02606}\ } (\bibinfo {year} {2017})}\BibitemShut
  {NoStop}%
\bibitem [{\citenamefont {Nixon}(1965)}]{FloydBook}%
  \BibitemOpen
  \bibfield  {author} {\bibinfo {author} {\bibfnamefont {F.~E.}\ \bibnamefont
  {Nixon}},\ }\href
  {https://books.google.es/books/about/Handbook_of_Laplace_transformation.html?id=QRsnAAAAMAAJ&redir_esc=y}
  {\emph {\bibinfo {title} {Handbook of Laplace transformation: fundamentals,
  applications, tables, and examples}}}\ (\bibinfo  {publisher}
  {Prentice-Hall},\ \bibinfo {year} {1965})\BibitemShut {NoStop}%
\bibitem [{\citenamefont {Feller}(1971)}]{FellerBook}%
  \BibitemOpen
  \bibfield  {author} {\bibinfo {author} {\bibfnamefont {W.}~\bibnamefont
  {Feller}},\ }\href
  {http://eu.wiley.com/WileyCDA/WileyTitle/productCd-0471257095.html} {\emph
  {\bibinfo {title} {An Introduction to Probability Theory and Its
  Applications}}}\ (\bibinfo  {publisher} {Wiley},\ \bibinfo {year}
  {1971})\BibitemShut {NoStop}%
\bibitem [{\citenamefont {Grabert}\ \emph {et~al.}(1987)\citenamefont
  {Grabert}, \citenamefont {Schramm},\ and\ \citenamefont
  {Ingold}}]{Grabert1985}%
  \BibitemOpen
  \bibfield  {author} {\bibinfo {author} {\bibfnamefont {H.}~\bibnamefont
  {Grabert}}, \bibinfo {author} {\bibfnamefont {P.}~\bibnamefont {Schramm}}, \
  and\ \bibinfo {author} {\bibfnamefont {G.-L.}\ \bibnamefont {Ingold}},\
  }\bibfield  {title} {\bibinfo {title} {\emph {Localization and anomalous
  diffusion of a damped quantum particle}},\ }\href {\doibase
  10.1103/PhysRevLett.58.1285} {\bibfield  {journal} {\bibinfo  {journal}
  {Phys. Rev. Lett.}\ }\textbf {\bibinfo {volume} {58}},\ \bibinfo {pages}
  {1285} (\bibinfo {year} {1987})}\BibitemShut {NoStop}%
\end{thebibliography}%
\end{document}